\documentclass[preprint,12pt]{elsarticle}

\usepackage{amssymb}
\usepackage{xcolor}
\usepackage{url} 
\usepackage{amsmath}

\journal{Optics \& Laser Technology}

\begin{document}

\begin{frontmatter}



\title{Shannon Entropy Helps Optimize the Performance of a Frequency-Multiplexed Extreme Learning Machine}


\author[1]{Marina Zajnulina} 

\affiliation[1]{organization={Multitel Innovation Centre},
            addressline={Rue Pierre et Marie Curie 2}, 
            city={Mons},
            postcode={7000}, 
            country={Belgium}}
\ead{marina@physik.tu-berlin.de; zajnulina@multitel.be}

\begin{abstract}
Knowing the dynamics of neuromorphic photonic schemes would allow their optimization for controlled data-processing capability in possibly simplified designs and minimized energy consumption levels. In nonlinear substrates such as optical fibers or semiconductors, these dynamics can widely vary depending on the encoded inputs, \textcolor{black}{even for a single set of physical parameters}. Thus, other approaches are required to optimize the schemes. Here, I consider a frequency-multiplexed Extreme Learning Machine (ELM) that encodes information in the line amplitudes of a frequency comb and processes this information in a single-mode fiber subject to Kerr nonlinearity. Its performance is evaluated with Iris and Breast Cancer Wisconsin classification datasets. I introduce the notions of Shannon entropy of optical power, phase, and spectrum and numerically show that the optimization of system parameters (continuous-wave laser optical power and the modulation depth of the subsequent phase modulator as well as the fiber group-velocity dispersion and length) yields the ELM performance that places this neuromorphic scheme among the top-performing state-of-the-art computer-based machine-learning models. I show that the ELM’s performance is robust against initial noise, paving the way for cost-effective designs. Using Soliton Radiation Beat Analysis, I show that information encoding symmetric in frequency-comb lines yields the formation of input-power-dependent Akhmediev-breather-like structures and Peregrine solitons, whereas asymmetric encoding of the comb exhibits an additional regime of soliton crystals. Also, I discuss that asymmetric encoding supports the theory of Four-Wave Mixing as a data processing mechanism, whereas symmetric encoding underlines the theory of soliton-mediated information processing. The findings advance the toolbox and knowledge of Neuromorphic Photonics and general Nonlinear Optics.
\end{abstract}

\begin{graphicalabstract}
\includegraphics[width= 0.95\textwidth]{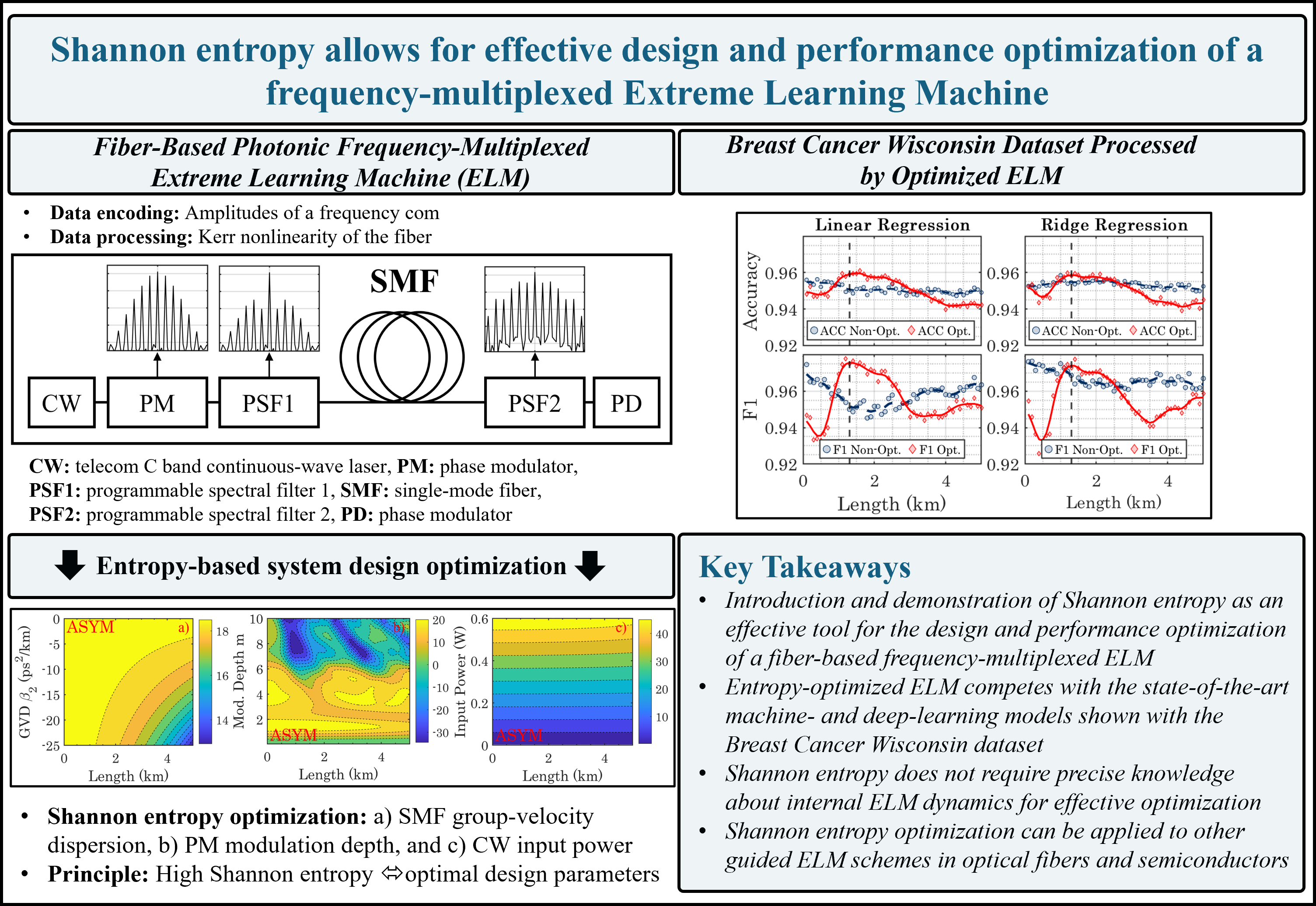}
\end{graphicalabstract}

\begin{highlights}

\item \textcolor{black}{Shannon entropy allows for effective ELM system design and performance optimization.}


\item \textcolor{black}{Entropy-optimized ELM competes with state-of-the art machine and deep learning models.}

\item \textcolor{black}{Symmetry of initial information encoding impacts data processing via FWM or solitons.}


\item \textcolor{black}{Symmetric encoding yields breathers and solitons; asymmetric also generates crystals.}

\item \textcolor{black}{ELM performance is robust to initial noise enabling cost-effective implementations.}

\end{highlights}


\end{frontmatter}



\section{Introduction}
\label{sec:INTRO}
Due to their ultra-wide bandwidths ranging up to tens of THz, comparably low energy consumption, and inherent analog nature, neuromorphic photonic schemes such as Optical Neural Networks (ONNs) constitute promising candidates to surpass the data-processing capabilities of the state-of-the-art electronic computing devices built upon von Neumann architecture. ONNs utilize light as an information carrier and various effects as information processing mechanisms \cite{McMahon2023}, \cite{Shastri2021}. Among other possibilities, optical data processing can be achieved by exploiting the Kerr effect in optical media \cite{Pauwels2019}, \cite{Fischer2023}, \cite{Zajnulina2023}, the nonlinearity of photo detectors \cite{Lupo2021} and built-in Mach-Zehnder modulators \cite{Brunner2018}, stimulated Brillouin scattering \cite{Phang2023}, optical mode interaction during multimode propagation \cite{Oguz2024}, spatial light modulation \cite{Yildrim2023} as well as light amplitude and phase changes due to passing through media \cite{Wetzstein2020}, Rayleigh scattering in optical fibers \cite{Redding2024}, \cite{Cox2024}, and linear wave scattering \cite{Wanjura2023}, \cite{Xia2024}. All these mechanisms have one feature in common: they implement some nonlinearity, either intrinsic to the optical medium, or being a characteristic of used optical components, or hidden in the waves scattering matrices. As in the case of computer-based artificial neural networks, implementing nonlinearity is a prerequisite for effective optical computing in ONNs. In addition, light offers the possibility of parallel data processing due to its multiple degrees of freedom, such as amplitude, frequency (or wavelength), phase, modes, and polarization \cite{McMahon2023}, \cite{Bai2023}, \cite{Xu2023_ONN}.

Due to their feasibility of implementation with optical and electro-optical components, two major types of ONNs are currently under study. The first type is the so-called Reservoir Computers that constitute neuromorphic counterparts of recurrent neural networks (cf. \cite{Pauwels2019}, \cite{Brunner2018}, \cite{Redding2024}, \cite{Argyris2018}, \cite{Duport2016}, \cite{Butschek2022}, \cite{Suzuki2022}). The second type is the Extreme Learning Machines (ELM) constituting feed-forward neural networks. In ELMs, the connections between the input and hidden layer remain untrained, only the connections between the hidden and output layer undergo a training procedure which often happens using linear models such as linear regression or ridge regression \cite{Huang2014}, \cite{Ding2014}, \cite{Duarte2023}. 

There exist a variety of successful propositions and implementations of photonic ELMs. Thus, Refs. \cite{Fischer2023}, \cite{Zhou2022} report continuous spectra for data encoding and subsequent transformation by Kerr nonlinearity in optical fibers. Ref.~\cite{Yildirim2023} utilizes low levels of second and third-order (Kerr) nonlinearity in a Lithium Niobate waveguide pumped with pulses of the order of tens of pJ. Ref.~\cite{Oguz2023} deploys the Kerr-nonlinearity-based coupling between the modes of \textcolor{black}{a} multi-mode fiber in their feed-forward neural network \textcolor{black}{(similar to an ELM) operating} at remarkable low-power consumption levels ($50~\text{nJ}$ pulses with an average power of $6.3~\text{mW}$). Ref.~\cite{Skontranis2023} reports a time-delayed ELM that utilizes a multimode Fabry-Perot laser as an accelerator. Ref.~\cite{Biasi2023} discusses an array of microresonators as an ELM, whereas Ref.~\cite{Pierangeli2021} presents a free-space ELM where spatial modulation of a laser beam is exploited to process data.

\textcolor{black}{Although nonlinearities are fundamental to the performance of ONNs, their internal dynamics have remained largely unexplored, and the question of how these dynamics correlate with or even determine data processing performance has only recently begun to attract attention (\cite{Fischer2023}, \cite{Zajnulina2023}, \cite{Saeed2025}, \cite{Ermolaev2025}, \cite{Sozos2024}, also cf. \cite{Morison_2024}, \cite{Sunada_2019}, \cite{Kesgin_2025}).} Rather, the ONNs have been treated as black boxes. \textcolor{black}{Yet this question is crucial, not only for a deeper understanding of the mechanisms underlying optical information processing but also for practical system optimization and control.} Thus, the answer might lead to a further decrease in power consumption and simplified system designs with improved data processing capabilities. \textcolor{black}{Also, it would contribute to the field of explainable AI (XAI) that aims at making data-processing systems transparent, interpretable, and accountable and, as a result, trustworthy.} 

\textcolor{black}{Admittedly, analyzing ONN dynamics is challenging: due to input-dependent encoding, a single set of physical parameters can give rise to a wide range of nonlinear regimes, varying across samples. This calls for practical, physically grounded methods capable of revealing relevant internal dynamics and guiding the targeted optimization of ONNs by allowing a map between their dynamics and data-processing capability. Also, despite the onset and growing interest to the internal dynamics and understanding of data-processing mechanisms in ONNS (\cite{Fischer2023}, \cite{Zajnulina2023}, \cite{Saeed2025}, \cite{Ermolaev2025}, \cite{Sozos2024}, \cite{Morison_2024}, \cite{Sunada_2019}, \cite{Kesgin_2025}), there are, to the best of my knowledge, no approaches available that have a generalization potential across different substrates and ONN schemes.} 

\textcolor{black}{Here, I introduce \textit{Shannon entropy} of optical power, phase, and spectrum and, using a frequency-multiplexed ELM as an example, show its strong potential to reveal ELM's internal dynamics allowing for its effective optimization. This handy approach requires comparably low computational resources and time and can be applied to other guided ELM schemes on optical-fiber and semiconductor substrates.}   

\textcolor{black}{The ELM I consider here utilizes} frequency multiplexing to encode the data and process them in a standard single-mode fiber (SMF) subject to Kerr nonlinearity (\cite{Zajnulina2023}). Light propagation in ELMs that utilize optical fibers as guiding and data-processing medium (cf. \cite{Fischer2023}, \cite{Zhou2022}, \cite{Oguz2023}) is described by the Nonlinear Schr\"{o}dinger Equation (NLS). Among other waves, this equation has solitons as possible solutions \cite{Agrawal2019}. In their theoretical paper (\cite{Marcucci2020}), the authors discuss that solitons play an important role in the training and computing of an ELM in a highly nonlinear regime (also cf. \cite{Bile2023}). Most recent studies (\cite{Ermolaev2025}, \cite{Saeed2025}) support this conclusion by experimentally showing a performance improvement with increasing soliton number for certain configurations and datasets in ELMs that use continuous spectra for information encoding and processing. In frequency-multiplexed ELMs that deploy frequency combs for information encoding and processing, it is rather Kerr-nonlinearity driven four-wave mixing (FWM) that facilitates computation as argued in Refs.~\cite{Zajnulina2023}, \cite{Sozos2024}. The authors of Ref.~\cite{Zajnulina2023} even show that a low level of Kerr nonlinearity is sufficient to effectively boost the performance of their ELM. In the anomalous-dispersion fiber regime (telecom C band), soliton formation and FWM happen in parallel, these two Kerr-nonlinearity driven effects are not to be kept apart with soliton formation counterbalancing FWM (\cite{Agrawal2019}). Therefore, despite the results presented in Ref.~\cite{Zajnulina2023}, the question of what mechanism,  FWM or soliton formation, prevails in information processing in a frequency-multiplexed ELM is not exhaustively answered yet.     

Apart from the fact that the internal ONN dynamics are generally complex due to the variance in data encoded and processed, a frequency-multiplexed ELM - as considered here - has an additional challenge. Namely, unmodulated and (information-)modulated frequency combs generate a plethora of various nonlinear waves, not only solitons, when they propagate through a Kerr medium such as an optical fiber. These nonlinear waves and their parameter spaces are not yet fully understood and constitute active research in the Nonlinear Optics community. Thus, various solitonic waves (single solitons and Akhmediev breathers), as well as their interactions and compounds in form of collisions, soliton molecules, crystals, and gas, have been reported in last years \cite{Finot2015}, \cite{Dudley2009}, \cite{Frisquet2013}, \cite{Xu2019}, \cite{Andral2020}, \cite{Schiek2021}, \cite{Zajnulina2024}, \cite{Suret2024}. In this perspective, the attempt to analyze and directly map the dynamics of a frequency-multiplexed ELM to its performance becomes even more challenging. \textcolor{black}{Yet again, there is a need for approaches that would provide us with insights about ELM's internal dynamics to use them for targeted ELM performance optimization}.

\textcolor{black}{As mentioned, I here introduce Shannon entropy as such an approach. A foundational version of this idea appears in Ref.~\cite{Yamano2024} where Shannon entropy is applied to the power of an NLS-governed field to derive conclusions about its soliton-related dynamics. Here, I extend and adapt this concept to match the specifics of a frequency-multiplexed ELM and further introduce new entropy measures for the optical phase and optical spectrum.} 
Using two bench classification tasks (Iris \cite{Fisher1936} and Breast Cancer Wisconsin \cite{Wisconsin1993} datasets), I numerically study the ELM performance and show that the notions of Shannon entropy of optical power, phase, and spectrum indeed allow for an effective optimization of ELM parameters such as optical input power, fiber group-velocity dispersion (GVD) parameter and length, the modulation depth of a deployed phase modulator, as well as the way of information encoding. With Breast Cancer Wisconsin dataset, I show that the optimized ELM takes the place among the state-of-the-art computer-based machine learning models, all this without the need to precisely know the internal ELM dynamics. In the case of the Iris dataset, I also provide an example of possible dynamical regimes that depend on the input power and information encoding type. For this, I use the so-called \textit{Soliton Radiation Beat Analysis} (SRBA)  that allows to retrieve soliton content of pulses in optical fibers that originate from arbitrary inputs \cite{Boehm_2006}, \cite{Zajnulina_2015}, \cite{Zajnulina_2017}, \cite{Mitschke2017}. 

Using Shannon entropy as an approach to indicate the internal dynamics of an ELM is justified as photonic neuromorphic schemes can be considered as high-dimensional nonlinear dynamical systems where Shannon entropy has been reported to be an effective dynamical indicator (cf. \cite {Cincotta2021}). In information theory, Shannon entropy is a measure of the information content of a system. Information, on the other hand, is a statistical notion of how unlikely is an event. Thus, unlike events coincide with high information. Generally, dynamical systems either conserve (constant entropy) or destroy information (decreasing entropy) \cite{Duran2022}. The entropy of a data-processing system can increase \cite{Kuyper1996}. Originally formulated via probabilities, Shannon entropy has recently extended to fields and their absolute values (powers) \cite{Yamano2024}, \cite{LIMA2022}.

I show that for the ELM fiber length considered here ($L \leq 5~\text{km}$) no precise description of the solitonic wave evolution can be given as the fiber is too short to have formed concrete solitonic waves for most chosen system parameters. I rather speak about proto-solitonic evolution. Using numerical simulations to determine ELM classification accuracy as well as the notions of Shannon entropy of optical power, phase, and spectrum, I show that these proto-solitonic regimes coincide with a fast increase in the ELM accuracy and highest values of entropies. The accuracy saturates or even declines with decreasing entropy values making the latter a suitable tool to optimize ELM's physical parameters (input power, fiber GVD parameter and length, modulation depth of the deployed phase modulator, and the type of information encoding). Due to the similarities of the ELM design, the introduced entropy notions would apply to systems presented in Refs.~\cite{Fischer2023}, \cite{Saeed2025}, \cite{Ermolaev2025}, \cite{Sozos2024}, and \textcolor{black}{potentially} other ELM schemes. Interestingly, decreasing entropy values coincide with the finalization of the solitonic-wave formation making the notion of Shannon entropy a potentially attractive tool for dynamics characterization in the Nonlinear-Optics community as well.

Having studied two types of information encoding in the amplitudes of the \textcolor{black}{initial} frequency comb, symmetric and asymmetric encoding, I show that asymmetric encoding leads to better ELM performance. As it also coincides with a higher degree of FWM, the results achieved with asymmetric encoding support the theory of FWM-driven information processing presented in Ref.~\cite{Zajnulina2023}. On the other hand, the results achieved with symmetric encoding being only negligibly minor to the asymmetric one support the theory of soliton-driven information processing presented in Ref.~\cite{Marcucci2020}. Using SRBA and a sample from the Iris dataset, I showed that symmetric information encoding would lead to input-power dependent formation of Akhmediev-breather-like structures and Peregrine solitons (being limit cases of Akhmediev breathers) and asymmetric encoding would induce input-power dependent transitions from Akhmediev-breather-like structures to soliton crystals and separated solitons if a longer fiber was deployed. These results contribute to a better understanding of the evolution of modulated frequency combs in fibers and are relevant for the Nonlinear-Optics community.

The paper is structured as follows: Sec.~\ref{sec:Setup} describes the ELM scheme under study, the methodology used for numerical simulations, introduces symmetric and asymmetric information encoding in the amplitudes of the lines of a frequency combs as well as the notions of Shannon entropies of optical power, phase, and spectrum. Sec.~\ref{sec:results} discusses ELM classification accuracy results achieved for different values of optical input power, GVD parameter, and the modulation depth of the phase modulator deployed to produce the initial comb. These results are linked to the corresponding Shannon entropy values to derive optimal system parameters. This is done for the Iris dataset and both information encoding types. Then, using the Breast Cancer Wisconsin dataset, I show the parameter optimization in action. Thus, I optimize the ELM using Shannon entropy of optical power and then compare the classification accuracy results of an optimized ELM with a non-optimized one showing the superiority of the optimized case. A conclusion is drawn in Sec.~\ref{sec:conclusion}. Sec.~\ref{sec:app1} presents the results and discussion of possible dynamical regimes achieved utilizing SRBA. 

\section{Methods}
\subsection{Extreme Learning Machine Setup}
\label{sec:Setup}

\begin{figure}[ht]
\centering
\fbox{\includegraphics[width= 0.60\textwidth]{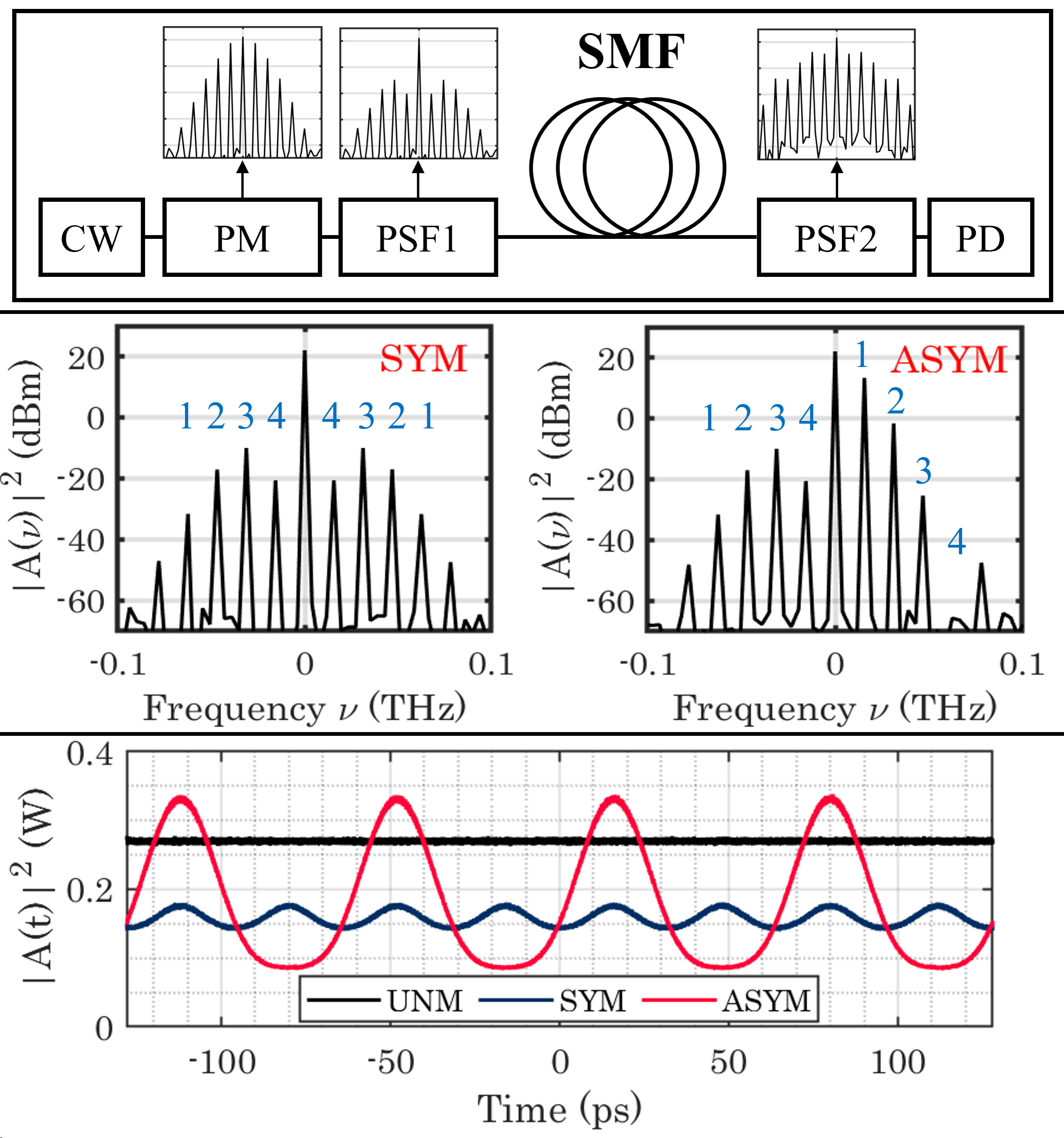}}
\caption{\textit{Top:} Experimental \textcolor{black}{scheme} of a frequency multiplexed Extreme Learning Machine with CW: continuous-wave laser, PM: phase modulator, PSF1: programmable spectral filter 1, SMF: standard single-mode fiber, PSF2: programmable spectral filter 2, and PD: photo diode \cite{Zajnulina2023}. \textit{Middle:} Symmetric (SYM) and asymmetric (ASYM) information encoding by PSF1 using the first sample of Iris dataset as an example. \textit{Bottom:} Comparison of the optical power of an unmodulated (UNM) initial comb and a comb modulated via SYM and ASYM for the first sample of Iris dataset and initial power of $P_{0} = 0.27~\text{W}.$}
\label{fig:SETUP_ENCODING}
\end{figure}

The schematic of a fiber-based frequency-multiplexed ELM studied here is shown in Fig.~\ref{fig:SETUP_ENCODING} (top). The ELM input layer consists of a continuous-wave laser (CW) in the telecom C band, a phase modulator (PM), and a programmable spectral filter (PSF1). The CW laser and the PM modulator generate an initial (unmodulated) frequency comb, in which PSF1 encodes information to process by attenuating the comb lines with values corresponding to the features of the dataset samples. The hidden layer is a standard single-mode fiber (SMF) in which the modulated comb is transformed under the impact of Kerr nonlinearity. The read-out layer consists of a second programmable spectral filter (PSF2) and a photodiode (PD) to collect the comb line intensities individually. The vectors of the comb line intensities are stuck into a matrix that is used for linear regression to generate the ELM output. The experimental realization and validation of a similar fiber-based frequency-multiplexed ELM are reported in Ref.~\cite{Zajnulina2023}. Contrary to the scheme depicted here (Fig.~\ref{fig:SETUP_ENCODING}, top), the authors of Ref.~\cite{Zajnulina2023} deploy an erbium-doped fiber amplifier (EDFA) between PSF1 and SMF to increase the output of PSF1 to invoke Kerr-nonlinear light propagation. They also show that the optical data processing primarily occurs in the EDFA they use rather than in the subsequent SMF stage. The internal dynamics of EDFAs are more complex than passive SMFs and include soliton compression effects due to amplification. To concentrate on the essential effects caused by fiber GVD and nonlinearity, I assume that the output of the PSF1 (Fig.~\ref{fig:SETUP_ENCODING}, top) is intense enough for nonlinear processes to take place such that no EDFA is needed.

\subsection{Modeling of Light Propagation in Single-Mode Fiber}
\label{sc:NLS}
The ELM is emulated on the computer by modeling its stages, and numerical simulations are used to generate the ELM output. For nonlinear light propagation in the SMF stage, I use NLS for the optical field amplitude $A(z,t)$ in the slowly varying envelope approximation in the co-moving frame \cite{Agrawal2019}:
\begin{equation}
\frac{\partial A}{\partial z} = -i\frac{\beta_2}{2}\frac{\partial^{2}A}{\partial t^{2}} + i\gamma |A|^{2}A 
\label{equ:NLS}
\end{equation}
with $\beta_2$ being the group-velocity dispersion (GVD) parameter and $\gamma$ the nonlinear coefficient at CW laser wavelength $\lambda_{0} = 1554.6~\text{nm}$ (cf. \cite{Zajnulina2023}). If not stated otherwise, $\gamma = 1.2~(\text{W}\cdot\text{km})^{-1}$ is used throughout the text, which is a typical value for standard SMFs (cf. Ref.~\cite{Zajnulina2023}). Optical losses, higher-order dispersion, Raman effect, and shock of short pulses usually present in standard fibers are omitted as they have a negligible impact for the SMF length of $L=5~\text{km}$ considered below.
The initial condition $A(z=0, t)$ provided by PM reads as: 
\begin{equation}
A(z=0, t) = \sqrt{P_{0}}\exp{\left(i\omega_{0} t + im \cos{(2\pi\Omega t)}\right)} + \sqrt{n_{\text{0/rand}}(t)}\exp{(i\phi_{\text{rand}}(t))}
\label{equ:IC}
\end{equation}
with PM modulation frequency $\Omega = 15.625~\text{GHz}$ and modulation depth $m.$ Rewriting the first term of Eq.~\ref{equ:IC} via Jacobi–Anger expansion shows the (unmodulated) comb nature of the initial condition: 
\begin{equation}
\sqrt{P_{0}}\exp{\left(i\omega_{0} t + im \cos{(2\pi\Omega t)}\right)} = \sqrt{P_{0}}\sum_{k=-\infty}^{+\infty}i^{k}J_{k}(m)\exp{(i\omega_{0}t + i k 2\pi\Omega t)},
\label{equ:JacobiAnger}
\end{equation}
where $J_{k}(m)$ is the $k-$th Bessel function of the first kind. The second term in Eq.~\ref{equ:IC} represents an additive initial white noise, with random amplitude and phase. If not stated otherwise (cf. Sect.~\ref{sec:noise}), I use the noise amplitude values that correspond to a spectral signal-to-noise ratio of \textcolor{black}{$\text{SNR} = 95~\text{dB}$}, denoting a low-noise system with a stable CW laser and a low-noise phase modulator PM (Fig.~\ref{fig:SETUP_ENCODING}, top).

The temporal window for simulations is chosen to be $256~\text{ps}$ raging from $t_{\text{start}} = -128~\text{ps}$ to $t_{\text{end}} = +128~\text{ps}.$ Numerical integration of Eq.~\ref{equ:NLS} occurs using the Fourth-Order Runge–Kutta in the Interaction Picture Method \cite{Hult2007}.

\subsection{Extreme Learning Machine Information Encoding}
\label{sec:encoding}
Below, I study the relationship between ELM performance and how information is encoded in initial comb lines produced by PM (Eq.~\ref{equ:IC}). I consider symmetric (SYM) and asymmetric (ASYM) encoding. To understand the difference between SYM and ASYM, let us consider the Iris dataset as an example. This dataset consists of 150 samples with 4 numerical features $X \in\mathbb{R}^{150 \times 4}$ each representing the characteristics of Iris flowers of 3 different classes such that the target vector is $y \in\mathbb{N}^{150 \times 1}$. The data set is balanced, i.e., there are 50 samples per class \cite{Fisher1936}.

First, the features of the $i-$th sample $\pmb{x}^{j}_{\text{dataset}}\in X$ undergo Min-Max normalization to be rescaled to the range $[0,1]:$
\begin{equation}
    \pmb{x}^{j} = \frac{\pmb{x}^{j}_{\text{dataset}}-\min(X)}{\max(X) - \min(X)}.
\label{equ:min_max}
\end{equation}

Then, each normalized input data sample $\pmb{x}^{j}$ is duplicated in the following way: 
\begin{equation*}
\pmb{x}^{j} = [x_{1}^{j}, x_{2}^{j}, x_{3}^{j}, x_{4}^{j}, 1, x_{4}^{j}, x_{3}^{j}, x_{2}^{j}, x_{1}^{j}]
\end{equation*}
for SYM and 
\begin{equation*}
\pmb{x}^{j} = [x_{1}^{j}, x_{2}^{j}, x_{3}^{j}, x_{4}^{j}, 1, x_{1}^{j}, x_{2}^{j}, x_{3}^{j}, x_{4}^{j}] 
\end{equation*}
for ASYM. 

Subsequently, the comb lines from $k= -4$ to $k = + 4$ (Eq.~\ref{equ:JacobiAnger}) are multiplied by $\pmb{x}^{j}$, which leads to their attenuation apart from the central line at the CW laser frequency $\omega_{0}$ (for convenience, $\omega_{0} = 0~\text{THz}$) as it is multiplied by 1. With this procedure, the symmetry of the modulated comb is preserved for SYM, whereas the symmetry is broken for ASYM (Fig.~\ref{fig:SETUP_ENCODING}, middle). In the time domain, the initially continuous input becomes pulsed after the application of SYM or ASYM encoding (Fig.~\ref{fig:SETUP_ENCODING}, bottom) \cite{Finot2015}. SYM and ASYM described here constitute specific examples of information encoding, other approaches of frequency-comb amplitude or comb-line phase modulation are possible (\cite{Lupo2021}, \cite{Zajnulina2023}). However, Ref.~\cite{Zajnulina2023} points out the superiority of comb-line amplitude modulation over phase modulation with respect to the performance of their fiber-based frequency-multiplexed ELM. Keeping this valuable result in mind, I concentrate on comb-line amplitude modulation only, \textcolor{black}{with} information encoding via phase modulation being skipped.

Tuning the PM modulation depth allows for the generation of combs with a various number of comb lines (Fig.~\ref{fig:SETUP_ENCODING}, top). However, this number is still limited, going maximally to a few tens. Thus, Ref.~\cite{Zajnulina2023} reports using a comb with 25 lines spanning a spectrum of $\approx3~\text{nm}$ for modulation depth $m\approx 2.$ This limitation might induce the impression that only simple datasets with a small number of features can be encoded in the ELM. Below, using the Wisconsin Breast Cancer dataset \cite{Wisconsin1993}, I show that a fiber-based ELM can also encode and efficiently process more complex datasets. This binary classification dataset comprises 569 samples with 30 features each, i.e. $X \in\mathbb{R}^{569 \times 30}.$ It is not balanced with 357 samples of benign and 212 samples of malignant tumors. Using Principal Component Analysis, I reduce the dimension of $X$ by choosing 4 principal components that account for $79.24\%$ of dataset's total variance, i.e. $X\in\mathbb{R}^{569 \times 30} \rightarrow \Tilde{X}\in\mathbb{R}^{569 \times 4}.$ Then, $\Tilde{X}$ goes through Min-Max normalization (Eq.~\ref{equ:min_max}) and is ASYM encoded in the comb lines in the same manner as the Iris dataset. It will be shown that the ELM performs in the range of state-of-the-art machine learning algorithms with a given number of principal components. Also, Ref.~\cite{Zajnulina2023} reports that weak Kerr nonlinearity is sufficient to effectively process data in a fiber-based ELM implying low input powers. Using this insight, I use input powers of $P_{0} \leq 0.6~\text{W}.$ 
 
\subsection{Data Read-Out and ELM Output Layer Training}
For each encoded data sample sent through the SMF for processing, I read out the amplitudes of 9 comb lines from $k= -4$ to $k = + 4$ positioned \textcolor{black}{around} $\omega_{0}.$ The corresponding vectors of 9 numerical values are stuck into a matrix that is subsequently used for linear, ridge, or logistic regression. 
A 2/3-part of the dataset is used to train the models and a 1/3-part to test them. A 100-fold cross-validation evaluates the ELM performance. Below, I consider two classification tasks, a multi-class (3 classes) classification with the Iris dataset \cite{Fisher1936} and a binary classification (2 classes) with the Breast Cancer Wisconsin dataset \cite{Wisconsin1993}. In the case of the Iris dataset, I use linear regression; \textcolor{black}{for} the Breast Cancer dataset, I compare \textcolor{black}{the} performance using linear, ridge, and logistic regression. For linear and ridge regression models, the target vectors of the classes are one-hot encoded, for logistic regression left as it is, i.e. as a vector of zeros and ones representing two different classes. To calculate the ELM outputs via linear, ridge, and logistic regression, I use the Python-based machine learning library \textit{scikit-learn} \cite{scikit-learn}. This step corresponds to the training of the ELM output layer \cite{Huang2014}, \cite{Ding2014}, \cite{Duarte2023}.

\subsection{Shannon Entropy for Extreme Learning Machine Optimization}
\label{sec:Shannon}
Following the definition of Shannon entropy of a field $A_{i}(z,t)$ evolving according to the NLS introduced in Ref.~\cite{Yamano2024}, I write: 
\begin{equation}
    E_{\text{pow}}(z) = -\frac{1}{G}\sum_{i=1}^{G}\int_{t_{\text{start}}}^{t_{\text{end}}} |A_{i}(z, t)|^{2}\log(|A_{i}(z, t)|^{2}/|A_{\text{unm}}(z = 0, t)|^{2})\text{dt}.
    \label{eq:Shannon_power}
\end{equation}
Here, the index $i$ denotes the $i-$th sample in the \textcolor{black}{subdataset drawn from the overall dataset (explanation follows)}. The samples are encoded either via SYM or ASYM encoding. The index $\text{unm}$ refers to the initial, unmodulated comb produced by PM (Fig.~\ref{fig:SETUP_ENCODING}, top). The corresponding optical power $|A_{\text{unm}}(z=0, t)|^{2}$  is used to normalize the arguments of the $\text{log}$ function. Thus, $E_{\text{pow}}$ has the unit of energy. \textcolor{black}{As the optical power $|A_{\text{unm}}(z=0, t)|^{2}$ is numerically treated as a vector, the division for normalization in the log-function argument is carried out pointwise. To ensure that there are no physical units within the function argument, other approaches such as normalization of the NLS (Eq.~\ref{equ:NLS}) itself are possible (\cite{Yamano2024})}.  

In the same manner, I define Shannon entropy of the optical phase:
\begin{equation}
    E_{\text{phase}}(z) = -\frac{1}{G}\sum_{i=1}^{G}\int_{t_{\text{start}}}^{t_{\text{end}}}\phi_{i}(z, t)^{2}\log(\phi_{i}(z, t)^{2}/\phi_{\text{unm}}(z=0, t)^{2})\text{dt}
    \label{eq:Shannon_phase}
\end{equation}
with phases 

\begin{equation*}
\phi_{i}(z,t) = \text{atan}2(\Im (A_{i}(z, t)) \textcolor{black}{,} \Re (A_{i}(z, t))) 
\end{equation*}
and 
\begin{equation*}
\phi_{\text{unm}}(z=0,t) = \text{atan}2(\Im (A_{\text{unm}}(z=0, t)) \textcolor{black}{,}  \Re (A_{\text{unm}}(z=0, t))).
\end{equation*}
For Shannon entropy of the optical spectrum, I write with Fourier transform $\mathcal{F}$:
\begin{equation}
    E_{\text{spec}}(z) = -\frac{1}{G}\sum_{i=1}^{G}\int_{s_\text{start}}^{s_\text{end}}|\mathcal{F}(A_{i})|^{2}\log(|\mathcal{F}(A_{i})|^{2}/|\mathcal{F}(A_{\text{unm}})|^{2})\text{d}\nu.
    \label{eq:Shannon_spec}
\end{equation}

There are several differences to the Shannon-entropy definition of Ref.~\cite{Yamano2024} that should be pointed out. 
Thus, contrary to the definition of Ref.~\cite{Yamano2024} that uses $t_{\text{start}} = -\infty$ and $t_{\text{end}} = +\infty,$ the integrals are evaluated between well-defined limits, i.e. $t_{\text{start}} = -128~\text{ps}$ and $t_{\text{end}} = +128~\text{ps}$ for Eqs.~\ref{eq:Shannon_power} and \ref{eq:Shannon_phase} as well as $s_{\text{start}}=-0.5~\text{THz}$ and $s_{\text{end}}= +0.5~\text{THz}$ for Eq.~\ref{eq:Shannon_spec}, which is due to the chosen temporal integration window of Eq.~\ref{equ:NLS} on the one hand and the restriction to the most informative spectral window in the Fourier space on the other hand. Further, to account for the richness of possible dynamics generated in the ELM by samples belonging to different classes, I average the entropy functions over $G$ samples drawn arbitrarily such that every class is represented by an equal number of samples. Thus, I chose $G=9$ for the Iris dataset and $G=6$ for the Breast Cancer Wisconsin dataset with 3 samples per class. The advantage of the sampling lies in the considerable minimization of the total calculation time as compared to the calculation of the entropies over the whole dataset.

\begin{figure}[ht]
\centering
\fbox{\includegraphics[width= 0.80\textwidth]{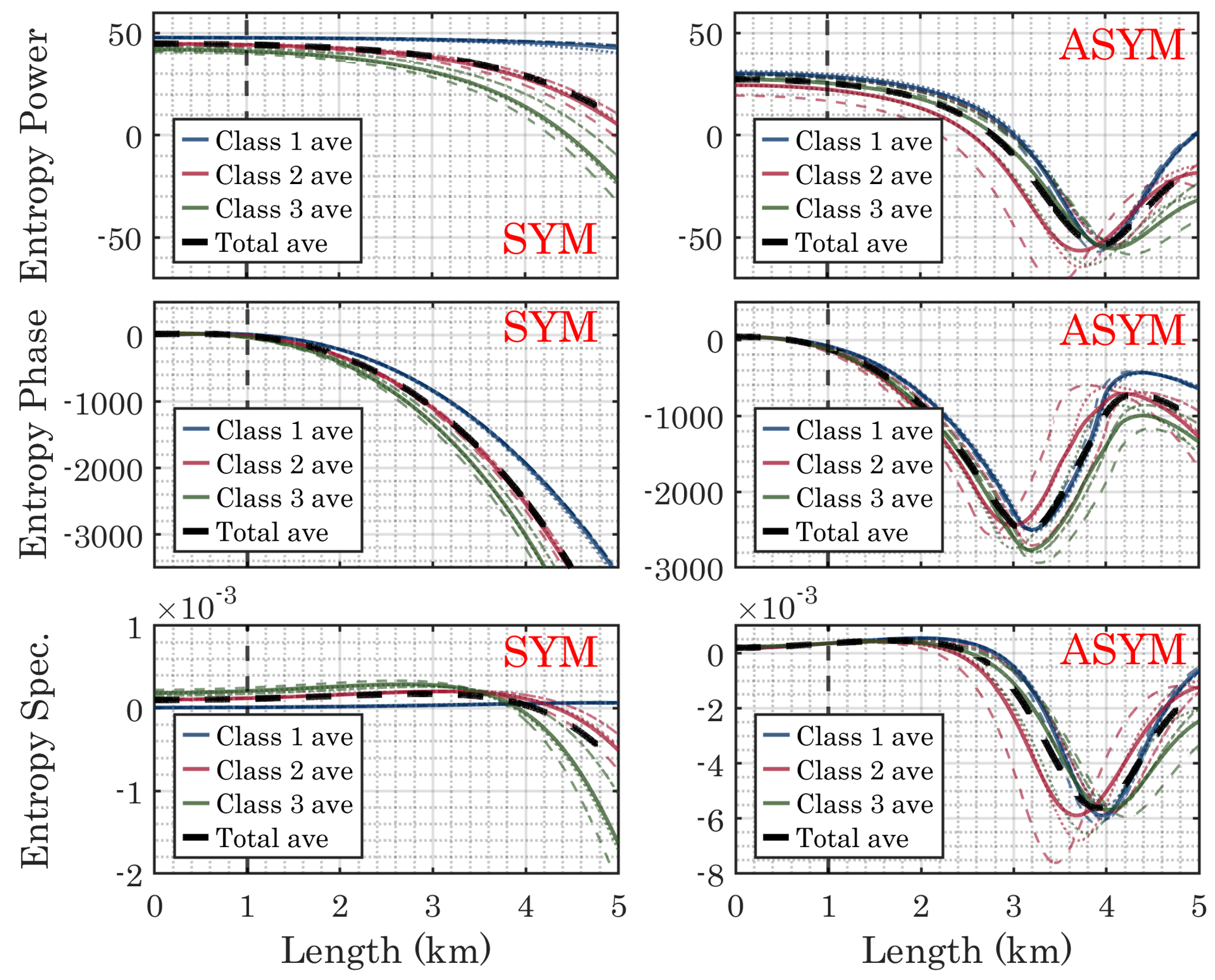}}
\caption{Shannon entropy for power (\textit{top}), phase (\textit{middle}), and spectrum (\textit{bottom}) for three arbitrary samples of each Iris dataset class, their averaged intra-class values (solid lines), and the average over all $G=9$ samples (black thick dashed line). \textcolor{black}{The SMF length is $L = 5~\text{km},$ GVD parameter $\beta_{2} = -23~\text{ps}^{2}/\text{km}$, PM modulation depth $m=1,$ and input power $P_{0}= 0.6~\text{W}.$}}
\label{fig:Shannon_AVE_Iris}
\end{figure}

Although the choice of 3 samples per class might seem small, it is sufficient to replicate the development of the entropy curves for the datasets analyzed here. Thus, Fig.~\ref{fig:Shannon_AVE_Iris} shows the examples of power, phase, and spectrum entropies (Eqs.~\ref{eq:Shannon_power}, \ref{eq:Shannon_phase}, and \ref{eq:Shannon_spec}, respectively) for $G=9$ samples of the Iris dataset, 3 samples per class. Different colors encode samples (dotted and dashed lines) from different classes. Solid colored lines average over the samples within a class, and the thick black dashed line averages over $G=9.$ As we can see, the evolution of the thick black dashed line represents well the evolution of the entropies of the samples and gives us a hint about the dynamics of the whole dataset which will be exploited below to optimize the ELM.    

\section{Results and Discussion}
\label{sec:results}
\subsection{A Note on the Dynamics} 
\label{sec:Dynamics_Note}
To consider the dynamics in the ELM SMF stage, let us first examine the output produced by encoding types SYM and ASYM. Fig.~\ref{fig:SETUP_ENCODING} (\textcolor{black}{bottom}) shows that both encoding types generate an input that is temporarily modulated in its optical power (cf. \cite{Finot2015}), whereby ASYM apparently produces stronger modulation of the initial CW power than SYM. As (anomalous) GVD is typically present in standard SMFs at CW laser wavelengths around $\lambda_0 = 1554.6~\text{nm}$ (telecom C band), the dynamics of the SYM and ASYM inputs are governed by the quasi-linear temporal Talbot effect at low input powers (\cite{Zajnulina2024}, \cite{Jannson_1981}). 

For higher input powers, both types of inputs, SYM and ASYM modulated, undergo modulational instability (MI) \cite{Agrawal2019}. MI enriches the spectrum by FWM and breaks the input into pulse trains compressing them into solitonic waves such as Akhmediev-breather-like structures (\cite{Dudley2009}, \cite{Frisquet2013}, \cite{Andral2020}), or soliton crystals, or separates solitons \cite{Zajnulina2024}, \cite{Zajnulina_2015}, \cite{Zajnulina_2017}, all depending on the input power and the depth of the CW modulation caused by information encoding. The input power needed to transit from quasi-linear temporal Talbot effect to the regime of solitonic-wave evolution can be estimated by the following relation:

\begin{equation}
    N^{2} := \frac{\gamma P_{0}}{(2\pi\cdot\Omega)^{2}|\beta_{2}|}
    \label{equ:Soliton_Order}
\end{equation}
with $N$ being the so-called soliton order \cite{Zajnulina_2017}. Solitonic evolution thresholds at $N = 0.5$ which leads to the value of $P_{0} = 0.046~\text{W}$ \cite{Boehm_2006}. It is important to note that this estimation is valid only for an unmodulated (UNM) comb injected into SMF. SYM and ASYM encoding in comb lines effectively reduce the optical power of the input which implies that encoded inputs require higher values of $P_{0}$ to transit \textcolor{black}{from a quasi-linear} to the nonlinear solitonic regime. 

In the solitonic regime, the propagation distance over which the first maximally compressed pulse train is observed decreases with increasing modulation depth \cite{Dudley2009}, \cite{Frisquet2013}, \cite{Andral2020}. For further distances, complex dynamics including breathers, separated solitons, soliton crystals and molecules, as well as separated-soliton gas is expected to evolve \cite{Dudley2009}, \cite{Frisquet2013}, \cite{Xu2019}, \cite{Andral2020}, \cite{Zajnulina2024}, \cite{Suret2024}. For an interested reader, a deeper-going discussion on possible dynamical regimes that can evolve in the SMF depending on the type of encoding (SYM/ASYM) and the level of the input power can be found in ~\ref{sec:app1}. It is important to note that these complex regimes will play hardly any role for considered SMF lengths of $L \leq 5~\text{km}$ as they usually evolve over much longer fiber lengths (several tens of kilometers).

The difficulty lies, however, in the fact that we still do not fully understand how a (modulated) comb as an input evolves to its first train of compressed pulses. Therefore, it is not yet possible to say with certainty what effect prevails in the first kilometers of the fiber, FWM or solitonic evolution, making the comparison and discussion of the results presented in Refs.~\cite{Zajnulina2024} and \cite{Marcucci2020} difficult, if not impossible. Fortunately, Shannon entropy, as proposed here (Sec.~\ref{sec:Shannon}) does not require precise knowledge of internal dynamics to allow for an effective ELM optimization.          

\begin{figure}[ht]
\centering
\fbox{\includegraphics[width= 0.98\textwidth]{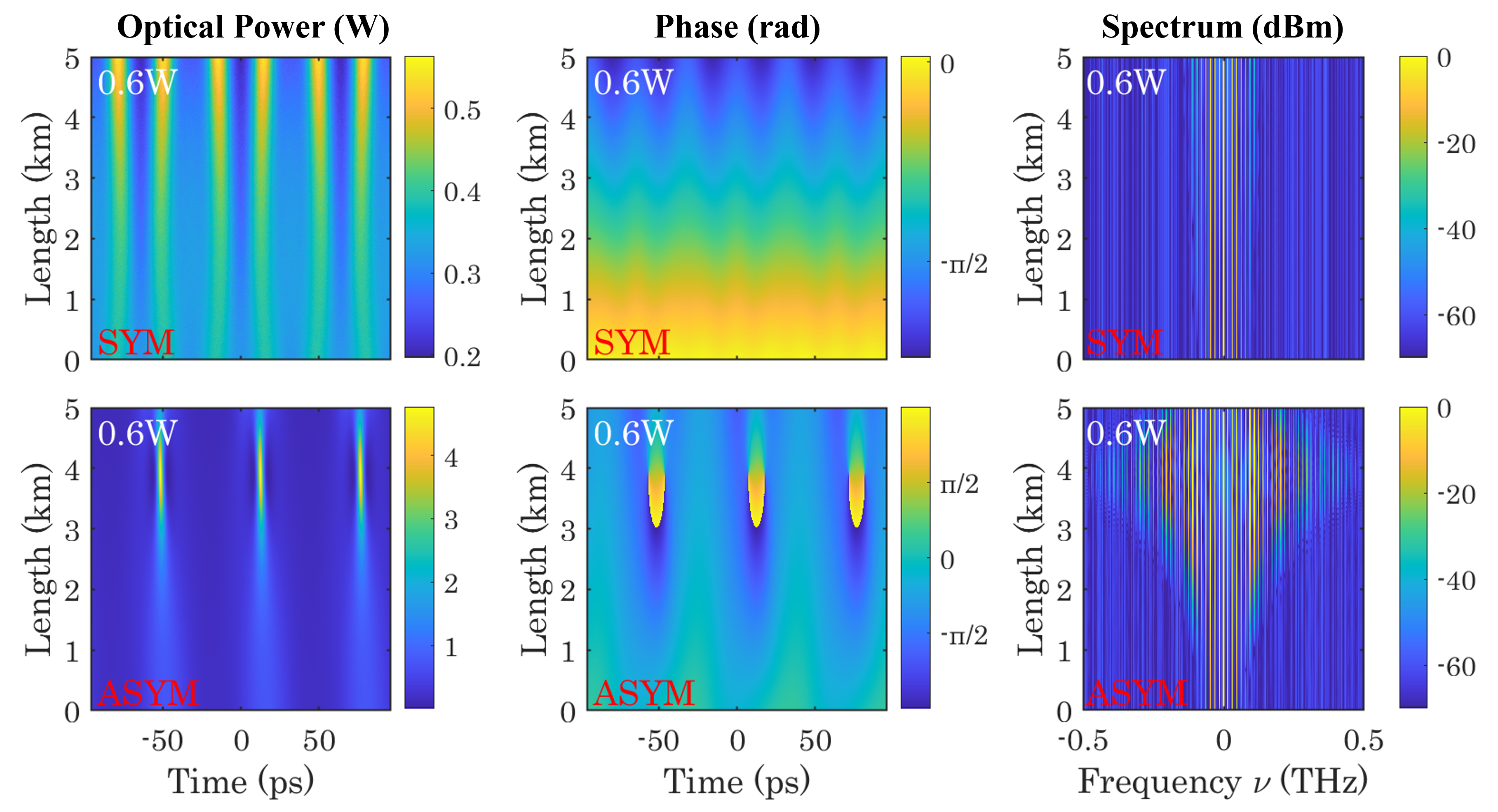}}
\caption{\textcolor{black}{Evolution of the optical power, phase, and spectrum for the first sample of the Iris dataset encoded symmetrically (SYM) and asymmetrically (ASYM) in the comb line amplitudes. The SMF length is $L = 5~\text{km},$ GVD parameter $\beta_{2} = -23~\text{ps}^{2}/\text{km}$, PM modulation depth $m=1,$ and input power $P_{0}= 0.6~\text{W}.$}}
\label{fig:BUILDUP}
\end{figure}

Let us now consider an example to gain an impression of how an information-encoded input evolves over the first kilometers of the SMF. Fig.~\ref{fig:BUILDUP} shows the evolution of the optical power, phase, and spectrum of the comb modulated via SYM and ASYM with the first sample from the Iris dataset in the SMF of length $L = 5~\text{km},$ GVD parameter $\beta_{2} = -23~\text{ps}^{2}/\text{km}$, PM modulation depth $m=1,$ and input power $P_{0}= 0.6~\text{W}.$ The sample is chosen arbitrarily (for simplicity, it is the first sample from the Iris dataset) and without loss of generality.    

For ASYM (Fig.~\ref{fig:BUILDUP}, bottom), we see a build-up of a train of compressed pulses at $L=4~\text{km}$. For SYM, such a build-up is not achieved for the considered SMF length. The maxima of the optical power drift in time with propagation length for ASYM which is caused by the asymmetry of the \textcolor{black}{information-encoded} frequency comb \cite{Schiek2021}. 

\begin{figure}[th]
\centering
\fbox{\includegraphics[width= 0.7\textwidth]{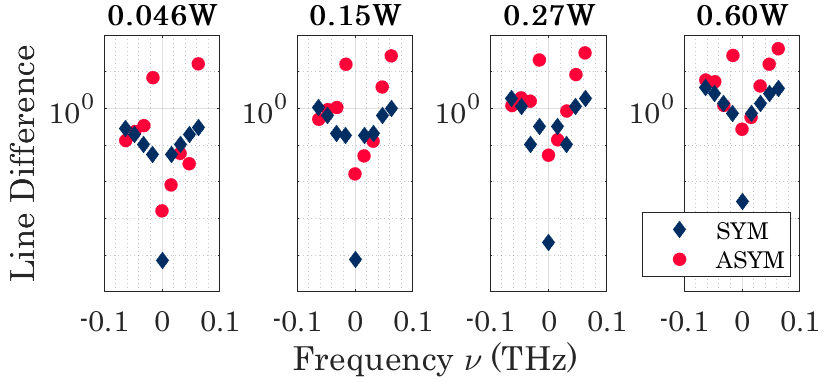}}
\caption{Difference between the amplitudes of SYM (blue) and ASYM (red) encoded lines at the input and output of the single-mode fiber with length of $L=5~\text{km}$ for different input powers $P_{0}$ and the first sample from Iris dataset as an example. The GVD parameter is $\beta_{2} = -23~\text{ps}^{2}/\text{km},$ the PM modulation depth is $m=1.$}
\label{fig:DIFF_lines}
\end{figure} 

When we consider the optical spectra, we see that the ASYM-modulated comb is subject to stronger FWM than the SYM-modulated one which results in a stronger transformation of frequency-comb lines amplitudes when they propagate through the SMF. We can see it in Fig.~\ref{fig:DIFF_lines} that depicts the difference in the amplitudes of the input and output frequency comb SYM and ASYM encoded with the first sample from the Iris dataset as an example. Thus, the ASYM encoded input undergoes a stronger change (higher values of line difference) in the SMF than the SYM input, specifically with increasing input power. According to the theory presented in Ref.~\cite{Zajnulina2023} that discusses FWM as the mechanism driving data processing in the ELM, an ELM with ASYM encoding should significantly outperform an ELM with SYM encoding. However, we will see that, although ASYM indeed leads to better ELM performance, the SYM-ELM output is still in the same order of magnitude, from which I conclude that (proto-)solitonic evolution (cf. Refs.~\cite{Marcucci2020}, ~\cite{Bile2023}) is also actively involved in data processing.

\subsection{Iris Dataset: Different Input Powers}
\label{subsec:iris_InputPower}

\begin{figure}[th]
\centering
\fbox{\includegraphics[width= 0.9\textwidth]{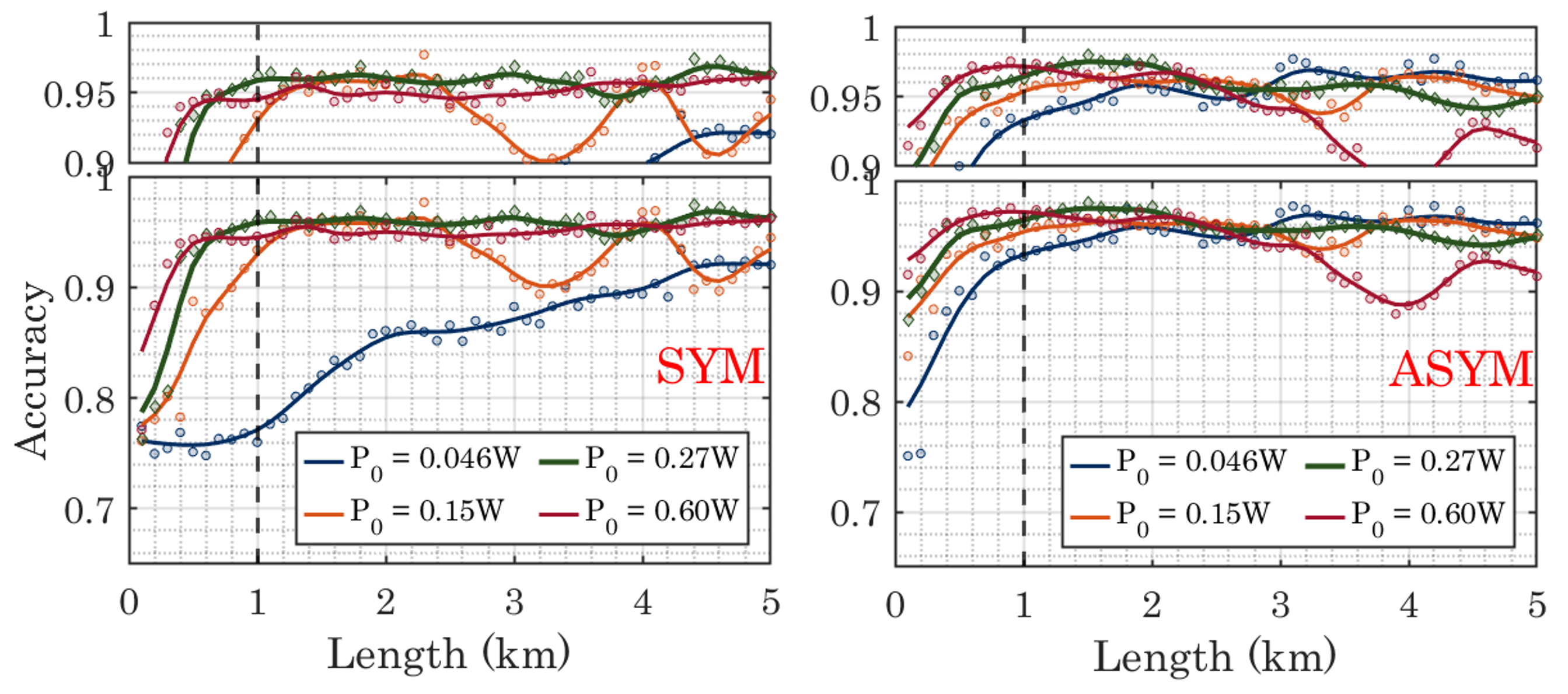}}
\caption{Classification accuracy for Iris dataset over the fiber length of $L = 5~\text{km}$ for symmetric (SYM) and asymmetric (ASYM) information encoding and different values of input power $P_{0},$ the PM modulation depth is $m=1.$ Upper panels represent the zoom-ins of the corresponding plots. The averaged standard deviation of the accuracy is $\approx\pm 0.03.$}
\label{fig:ACC_Length}
\end{figure} 

Now, we can proceed with studies of ELM performance and optimization depending on different system parameters. Fig.~\ref{fig:ACC_Length} shows Iris-dataset accuracy evolution in the ELM over an SMF section of $L= 5~\text{km}$ for SYM and ASYM at different input powers $P_{0}$ (Eq.~\ref{equ:IC}). The GVD parameter is $\beta_2 = -23~\text{ps}^{2}/\text{km}$ and the PM modulation depth $m=1.$ 

In general, for SMF lengths of $L\leq 1\text{km},$ we see a rapid increase of classification accuracy with increasing input power for both types of information encoding. The accuracy curves increase the faster, the higher \textcolor{black}{the} input power value $P_{0}$ (also cf. \cite{Saeed2025}, \cite{Ermolaev2025}). For longer lengths, the ELM accuracy performance is driven by the dynamics that depend on the input power and type of information encoding. In general, the ASYM-encoded ELM performs better than the SYM-encoded one achieving slightly higher accuracy values for all considered input powers. Thus, the maximum accuracy value achieved with ASYM is $0.98,$ whereas $0.97$ with SYM.

For $P_{0} = 0.046~\text{W},$ we see that the SYM case is inferior to ASYM with its slow increase of the accuracy curve and never reaching the accuracies at other values of $P_{0}$ for the considered SMF length. The ASYM accuracy curve reaches an accuracy value that is comparable to its counterparts after the propagation length of $L = 2.2~\text{km}.$ For this input power, the ELM operates in the dynamic regime that is governed by the quasi-linear Talbot effect (\ref{sec:app1}) for both, SYM and ASYM. Here, the impact of Kerr nonlinearity is negligible meaning that the data processing occurs mainly in the linear regime. This result is counterintuitive as a nonlinearity is needed to effectively process data in neuromorphic schemes. This nonlinearity might be hidden in the Talbot-effect-based scattering mechanisms of the optical power (Fig.~\ref{fig:OPT_DYNAMCIS}, cf. \cite{Wanjura2023}). As this observation opens the door for low-input-power on-chip integration of a frequency-multiplexed ELM, it needs further studies. Also, further studies are needed on the quasi-linear and nonlinear temporal Talbot effect with frequency-modulated inputs as we still lack a deeper understanding. 

For $P_{0} = 0.15~\text{W},$ the SYM-encoded ELM operates in the regime that goes towards the evolution of Akhmediev-breather-like structures, whereas the ASYM-encoded ELM is in the regime of developing soliton crystals and separated solitons (\ref{sec:app1}). As we can see, the SYM curve starts oscillating at a propagation length of $L>2.5~\text{W},$ whereas the ASYM curve evolves more smoothly. This indicates that soliton crystals might be more suitable for frequency-multiplexed ELMs due to the higher robustness as compared to Akhmediev breathers which is discussed in Refs. ~\cite{Zajnulina_2015}, \cite{Zajnulina_2017}.   

For $P_{0} = 0.27~\text{W}$ and $P_{0} = 0.6~\text{W},$ the ELM operates in the regime that goes towards the development of separated solitons for both types of information encoding \textcolor{black}{(}\ref{sec:app1}\textcolor{black}{)}. In these regimes, the accuracy curves quickly achieve their maximal values. Whereas the accuracy curves are smooth for SYM, the ASYM accuracy starts oscillating for SMF lengths $L>3~\text{km}.$ This oscillatory behavior is seen for $P_{0} = 0.6~\text{W}$ with its dip corresponding to the formation of a train of compressed pulses with a broad spectrum (cf. Fig.~\ref{fig:BUILDUP}). For both types of information encoding, we see that it is proto-soliton formation (\ref{sec:app1}) that drives information processing in the ELM supporting the findings of Ref.~\cite{Marcucci2020}. On the other hand, FWM as an information processing mechanism (\cite{Zajnulina2023}) can be even detrimental if it is too strong as it is the case with compressed pulses for ASYM, $P_{0} = 0.6~\text{W}.$   

\begin{figure}[th]
\centering
\fbox{\includegraphics[width= 0.9\textwidth]{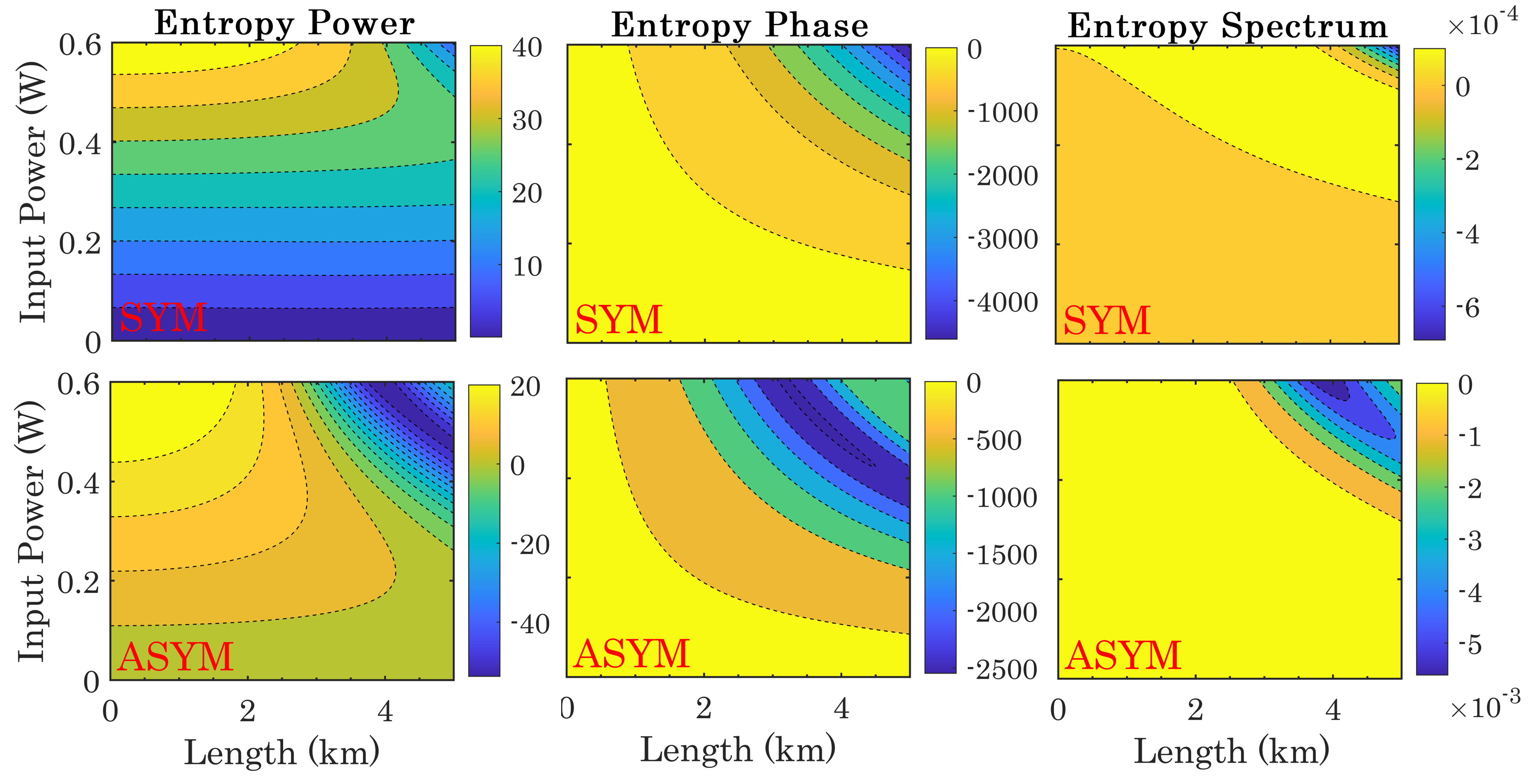}}\caption{Shannon entropy of optical power, phase, and spectrum for different input powers $P_0$ and information encoding types SYM and ASYM averaged over $G = 9$ arbitrary samples of the Iris dataset. The SMF length is $L = 5~\text{km},$ GVD parameter $\beta_{2}=-23~\text{ps}^2/\text{km},$ and PM modulation depth $m = 1.$}
\label{fig:Entropy_Power_Iris}
\end{figure}

To conclude, the SMF length should be kept short (I choose  $L=1~\text{km}$ as an optimum length) to avoid oscillatory behavior of the ELM accuracy over the propagation distance in the nonlinear regime. The ELM performance in the regime of the quasi-linear temporal Talbot effect still needs further studies. Now, let us take a look at the evolution of Shannon entropies (Eqs. \ref{eq:Shannon_power}, \ref{eq:Shannon_phase}, \ref{eq:Shannon_spec}) presented in Fig.~\ref{fig:Entropy_Power_Iris}.

The entropy of optical power (Fig.~\ref{fig:Entropy_Power_Iris}, left column) increases with input power towards its maximum in yellow in SMF lengths $L<2~\text{km}$ for SYM and ASYM. Specifically for $L<1~\text{km},$ a comparison with Fig.~\ref{fig:ACC_Length} reveals that higher entropy denotes a faster increase in accuracy. For fixed input powers, a decrease in entropy over the fiber length coincides with a decrease in the ELM classification performance. In particular, it is well seen for ASYM at $P_0 = 0.6~\text{W},$ where the drop in power entropy goes along with a dip in the ELM classification accuracy at $L= 4~\text{km}.$ However, the oscillation of the SYM-encoded accuracy curve for $P_{0} = 0.15~\text{W}$ does not directly transfer to the Shannon entropy of optical power. Thus, we do not see any dips in the power entropy. This is probably because the calculated entropy utilizes only $G = 9$ samples whereas the accuracy curves were produced with all 150 samples of the Iris dataset. It implies that with an increased number of samples used to calculate Shannon entropy, a better mapping between the accuracy curve and the entropy should become visible. However, even with $G=9,$ the Shannon entropy of optical power represents a good tool for the optimization of the ELM if we target the highest entropy values (i.e. yellow or yellowish color in entropy plots). Thus, the highest power entropy values are achieved for input powers $P_0 = 0.45 - 0.6~\text{W}$ and SMF lengths $L<2.2 ~\text{km}$ for SYM and $P_0 = 0.45 - 0.6~\text{W}$ and SMF lengths $L <2.1~\text{km}$ for ASYM. 

Shannon entropy of the optical phase  (Fig.~\ref{fig:Entropy_Power_Iris}, middle column) gives us a better orientation of what parameter space is to avoid for better ELM classification performance. Here, we also want to keep the entropy as high as possible as its high values correspond to a high increase and high values of ELM accuracy performance, whereas its decrease coincides with a stagnation or even a decrease in the accuracy curves (cf. Fig.~\ref{fig:ACC_Length}). Thus, according to the phase entropy plots, the optimal SMF length should be decreased to $L = 0.6 - 1~\text{km}$ for $P_{0}\rightarrow 0.6~\text{W}$ and both types of information encoding.  

Shannon entropy of the spectrum (Fig.~\ref{fig:Entropy_Power_Iris}, right column) provides us with an insight into the parameter space of the maximal spectral broadening of the input over the fiber length. Those are the regions of decreased entropy (blue) where the process of FWM is slow-downed and, at the point of maximal \textcolor{black}{temporal pulse} compression, counterbalanced by the dispersion. Ref.~\cite{Zajnulina2023} points out the importance of FWM for data processing in the ELM. Thus, parameter-space regions of slowed-down FWM and, accordingly, reduced spectrum entropy are to be avoided. In the context of this study, Shannon entropy of the spectrum does not add any value to the knowledge gained by considering the entropies of optical power and space. Therefore, I will omit it in the following discussion. However, in an experiment, spectral entropy would be easier to measure than phase entropy. Thus, along with the optical power entropy, it could be a preferred quantity for optimization studies. Also, this is an interesting function to study frequency comb generation and evolution in a broader context of \textcolor{black}{Nonlinear Optics.}

\subsection{Iris Dataset: Different Group Velocity Dispersion Parameters}
\label{subsec:iris_DiffBeta2}

\begin{figure}[bht]
\centering
\fbox{\includegraphics[width= 0.9\textwidth]{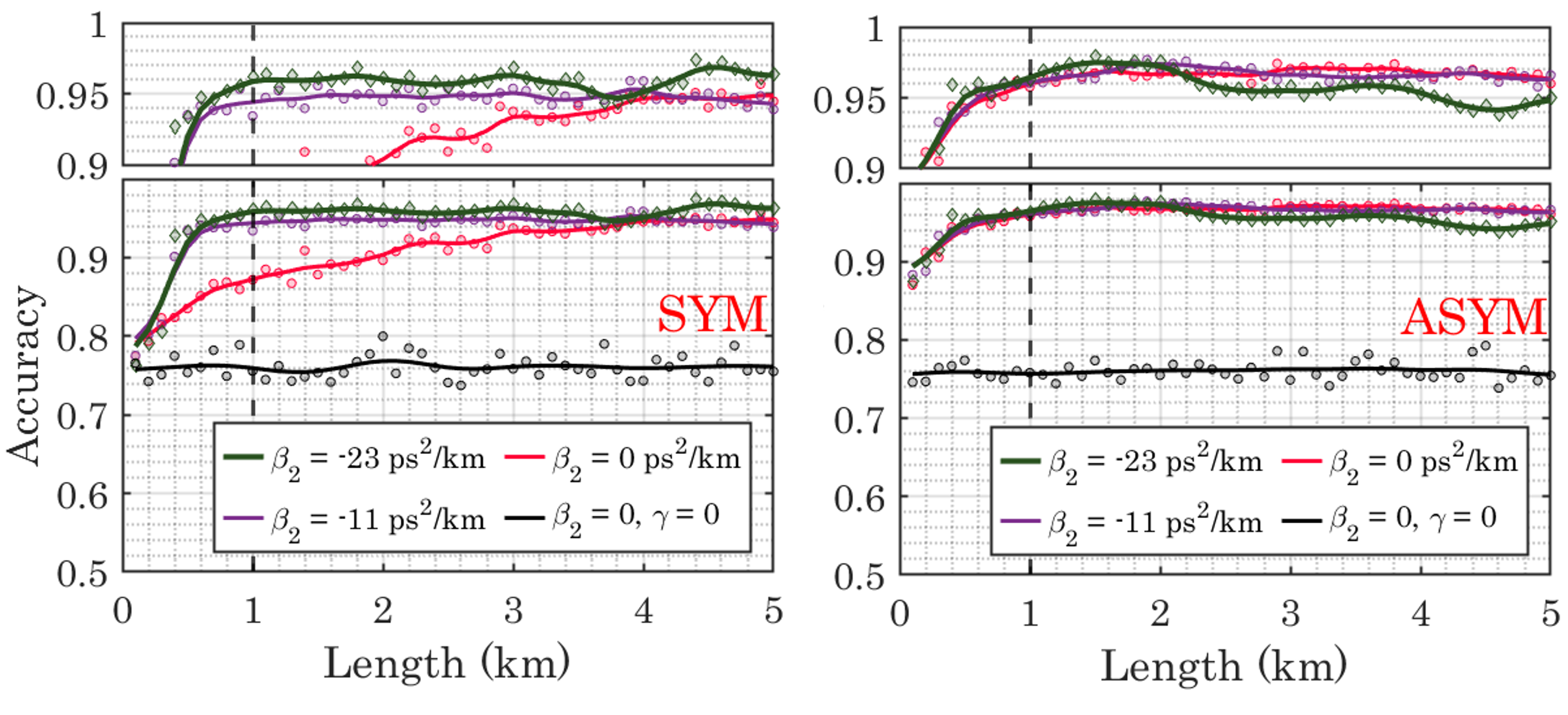}}
\caption{Classification accuracy for Iris dataset over the fiber length of $L = 5~\text{km}$ for symmetric (SYM) and asymmetric (ASYM) information encoding, input power $P_{0}=0.27~\text{W},$ different values of GVD parameter $\beta_{2},$ and PM modulation depth of $m=1.$ The black curves serve as a reference for worst possible ELM performance. Upper panels represent the zoom-ins of the corresponding plots. The averaged standard deviation of the accuracy is $\approx\pm 0.03$ for $\gamma = 1.2~(\text{W}\cdot\text{km})^{-1}$ and $\approx\pm 0.053$ for $\gamma = 0~(\text{W}\cdot\text{km})^{-1}.$}    
\label{fig:DIFF_Beta2}
\end{figure}

Fig.~\ref{fig:DIFF_Beta2} shows Iris-dataset accuracy evolution in the ELM over an SMF section of $L= 5~\text{km}$ for SYM and ASYM for different values of the GVD parameter $\beta_{2}$ at a fixed input power of $P_{0} = 0.27~\text{W}$ (Eq.~\ref{equ:IC}), the PM modulation depth is $m=1.$ For SYM, accuracy curves increase with the absolute value of $\beta_2$ which supports solitonic-waves-based information processing in the ELM (\cite{Marcucci2020}, \cite{Saeed2025}), rather than the FWM theory (\cite{Zajnulina2023}). Otherwise, the increase of the accuracy curves would be slowed down with increasing absolute value of GVD as dispersion counterbalances the effect of FWM \cite{Agrawal2019}. On the other hand, there is almost no dependence of the accuracy on the value of $\beta_2$ for ASYM which supports the theory of FWM driving the data processing in the ELM for this specific type of encoding (cf. ~\cite{Zajnulina2023}, \cite{Sozos2024}). In both cases, solitonic evolution goes along with FWM as soon as $\beta_{2}<0,$ and these two effects are not separable. Both of them contribute to the data processing in the ELM, with ASYM leading to a stronger contribution of FWM than SYM. 

Black curves $(\gamma = 0~(\text{W}\cdot\text{km})^{-1})$ depict the worst possible ELM performance and show that the Kerr nonlinearity is needed for the ELM to perform as a data-processing unit for the chosen input power. This result contradicts the observation of the SYM and ASYM accuracy increase at $P_{0}=0.046~\text{W}$ (Fig.~\ref{fig:ACC_Length}) that happens in the regime of a quasi-linear temporal Talbot effect. Whereas some nonlinearity is needed for implementation of neural networks in general, and the Kerr nonlinearity is the natural choice in optical guided Reservoir and Extreme Learning Machine schemes (\cite{Pauwels2019}, \cite{Fischer2023}, \cite{Zajnulina2023}, \cite{Saeed2025}), their data-processing capability differences in the quasi-linear and nonlinear regime certainly need further studies as they would allow for material and power consumption optimization opening the door for on-chip integration with a potentially reduced setup complexity.

\begin{figure}[bht]
\centering
\fbox{\includegraphics[width= 0.6\textwidth]{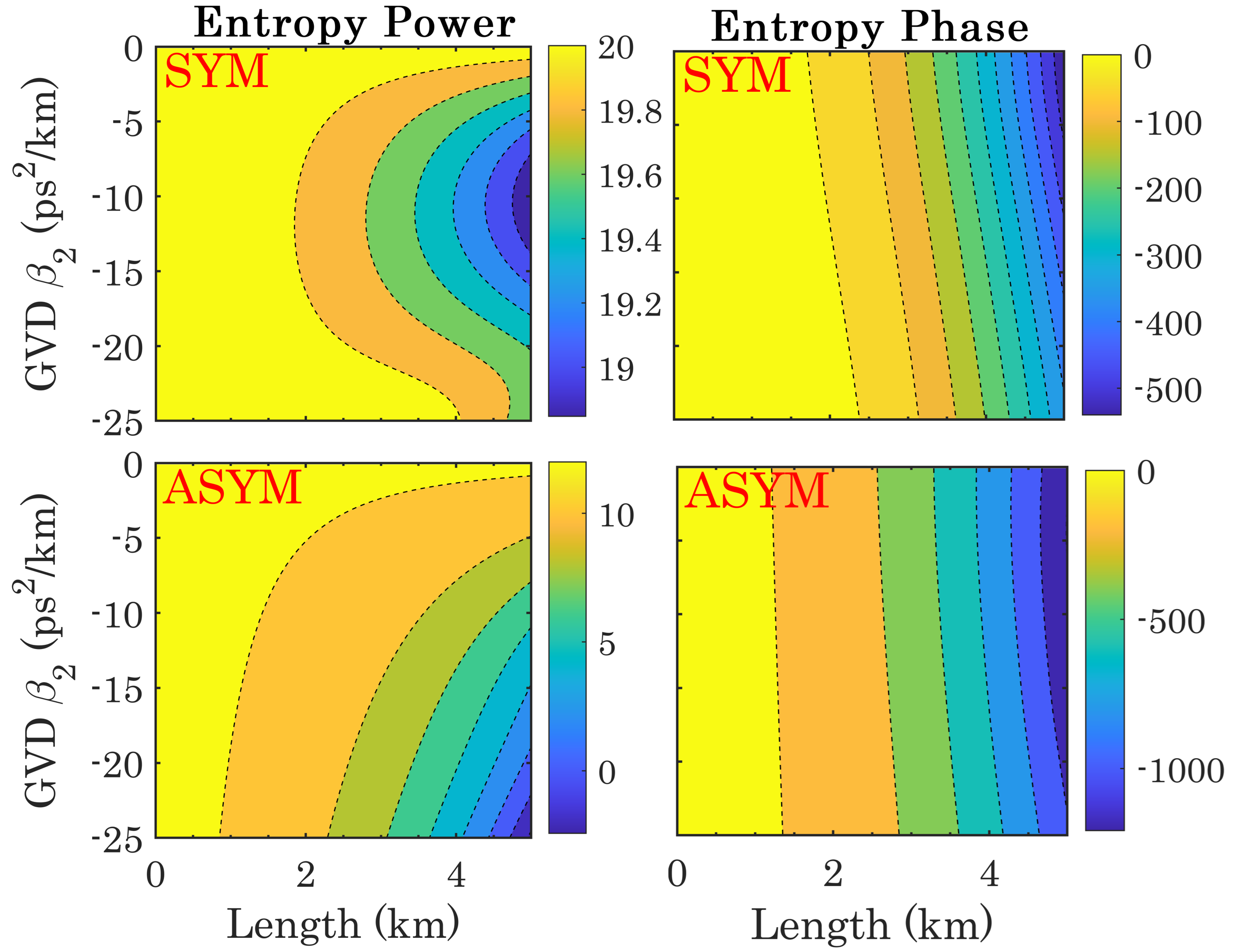}}
\caption{Shannon entropy of optical power and phase for different values of the GVD parameter $\beta_{2}$ and information encoding types SYM and ASYM averaged over $G = 9$ arbitrary samples of the Iris dataset. The SMF length is $L = 5~\text{km},$ input power $P_{0}= 0.27~\text{W},$ and PM modulation depth $m = 1.$}    
\label{fig:Entropy_Beta2_Iris}
\end{figure}

Fig.~\ref{fig:Entropy_Beta2_Iris} shows the corresponding Shannon entropy of optical power and phase for GVD parameter range of $\beta_{2}= -25~\text{ps}^2/\text{km}$ to $\beta_{2}=0~\text{ps}^2/\text{km}$ and input power $P_{0} = 0.27~\text{W}.$ Comparing Fig.~\ref{fig:Entropy_Beta2_Iris} with Fig.~\ref{fig:DIFF_Beta2}, we again can see that the highest values of entropies (yellow) correspond to the fast increase and/or highest values of the accuracy curves whereas a decrease in the entropies coincides with the stagnation or worsening of the accuracy performance giving us a tool to optimize the ELM with respect to the fiber dispersion. For both types of information encoding, SYM and ASYM, it is preferable to use low-dispersion fibers. This type of fibers is usually more expensive than standard SMFs, the cost factor could be a further optimization aspect. Standard SMFs, as considered here, exhibit a GVD dispersion of $\beta_{2}= -15~\text{ps}^2/\text{km}$ to $\beta_{2}=-25~\text{ps}^2/\text{km}$ around $\lambda_{0} = 1.55~\text{nm}.$ According to Fig.~\ref{fig:Entropy_Beta2_Iris}, these GVD values fit an optimized scheme if the SMF length is kept short, i.e. $L<1.2~\text{km}$ to account for both, SYM and ASYM encoding. For simplicity, I will proceed with the GVD parameter of $\beta_{2}=-23~\text{ps}^2/\text{km}.$ 

\subsection{Iris Dataset: Different Values of Phase-Modulator Modulation Depth}
\begin{figure}[bht]
\centering
\fbox{\includegraphics[width= 0.9\textwidth]{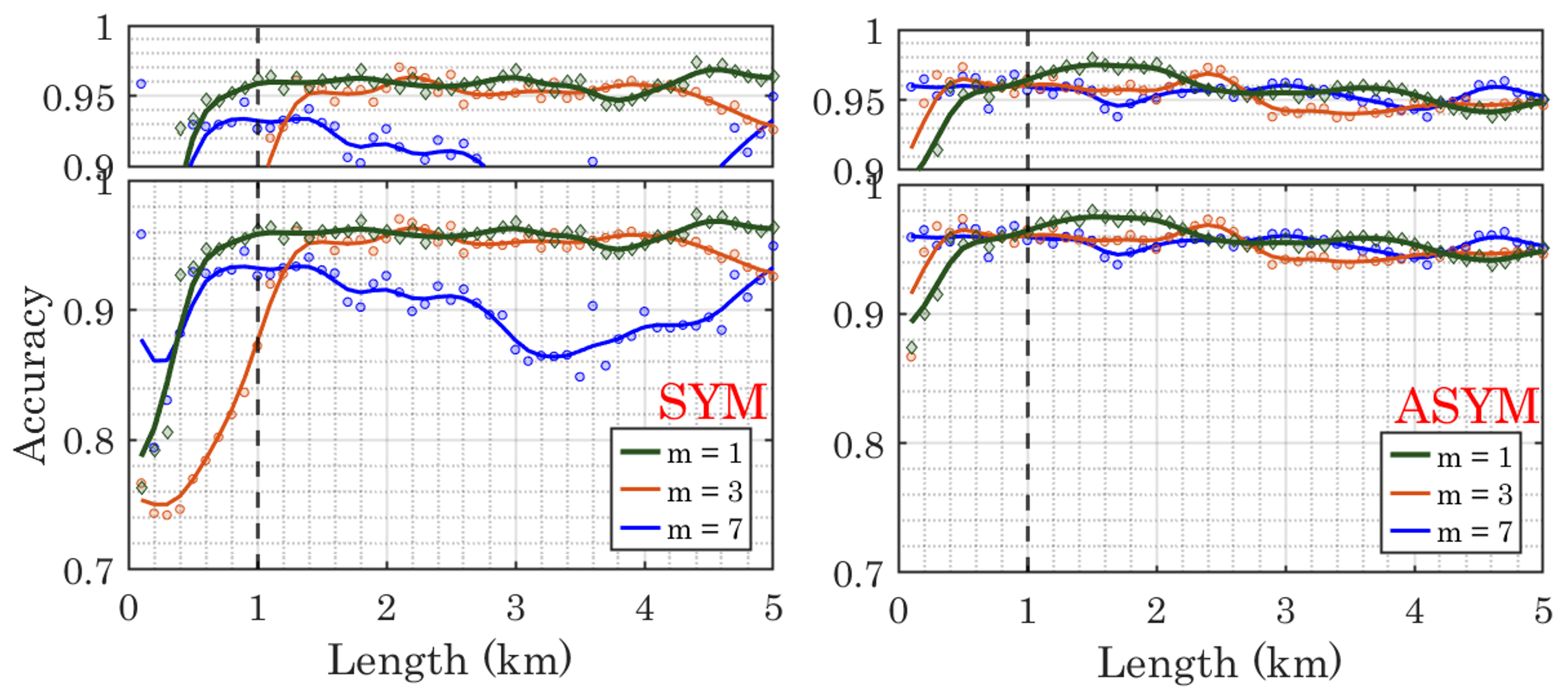}}
\caption{Classification accuracy for Iris dataset over the fiber length of $L = 5~\text{km}$ for symmetric (SYM) and asymmetric (ASYM) information encoding, input power $P_{0}=0.27~\text{W},$ GVD parameter of $\beta_{2} = -23~\text{ps}^{2}/\text{km},$ and different values of PM modulation depth $m.$ The averaged standard deviation of the accuracy is $\approx\pm 0.035$ for SYM and for $\approx\pm 0.027$ for ASYM.}    
\label{fig:DIFF_m}
\end{figure}
The modulation depth $m$ of the phase modulator PM in Fig.~\ref{fig:SETUP_ENCODING} (top) determines the number of lines in the initial (UNM) frequency comb: the higher the value of $m,$ the more lines the UNM comb has. This is an important parameter as it sets a limitation for the datasets that can be processed by the ELM. Thus, the natural assumption is to use a high value of $m$ for datasets that have a high number of features to encode. So far, I used $m=1.$ It generated a comb with 11 lines which was sufficient to encode the Iris dataset via SYM and ASYM. Still using this dataset, I now want to analyze whether an increase of $m$ affects the ELM performance. Indeed, it does as seen in Fig.~\ref{fig:DIFF_m} produced for $P_{0}= 0.27~\text{W},$ $\beta_{2}=-23~\text{ps}^2/\text{km}$ and modulation-depth values of $m = 1, 3, 7.$ Whereas the ASYM-ELM performance is seemingly independent of the value of $m$ (with a small exception of $m=1$ that shows slightly better performance than the rest), the performance of the SYM-ELM case suffers with the increase of $m.$

\begin{figure}[bht]
\centering
\fbox{\includegraphics[width= 0.6\textwidth]{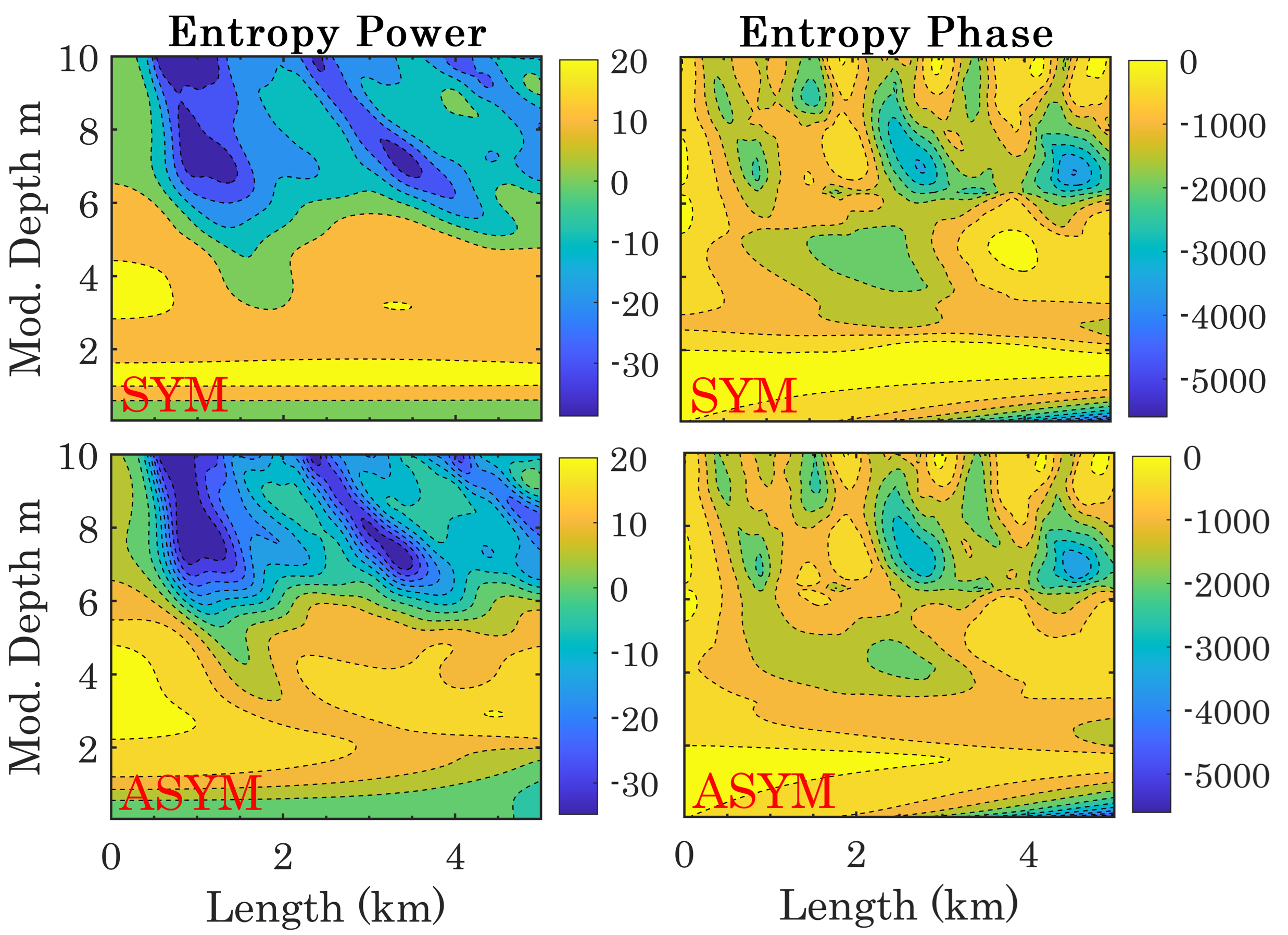}}
\caption{Shannon entropy of optical power and phase for different values of PM modulation depth $m$ and information encoding types SYM and ASYM averaged over $G = 9$ arbitrary samples of the Iris dataset. The SMF length is $L = 5~\text{km},$ input power $P_{0}= 0.27~\text{W},$ and GVD parameter $\beta_{2}= -23~\text{ps}^{2}/\text{km}.$
}    \label{fig:Entropy_m_Iris}
\end{figure}

Shannon entropies of optical power and phase produced for various values of $m$ (Fig.~\ref{fig:Entropy_m_Iris}) indicate complex dynamics evolution for $m>4.5$ for both, SYM and ASYM. For $m<4.5,$ ASYM, however, shows bigger parameter regions with higher entropy which explains its alsmost independent behavior with respect to the value of $m$ (Fig.~\ref{fig:DIFF_m}). To account for both types of encoding, low modulation depth values ($m<3$) should be chosen for the implementation of an ELM. If the number of the provided comb lines is too low with respect to the dataset's feature space, dimensionality reduction techniques should be applied to make sure that the number of encodable features is compatible with the number of available comb lines. Below, I will show that Principal Component Analysis (PCA) is a suitable approach to reduce the dimension of the features to encode. 

\subsection{Iris Dataset: Different Initial Noise Levels}
\label{sec:noise}

\begin{figure}[bht]
\centering
\fbox{\includegraphics[width= 0.98\textwidth]{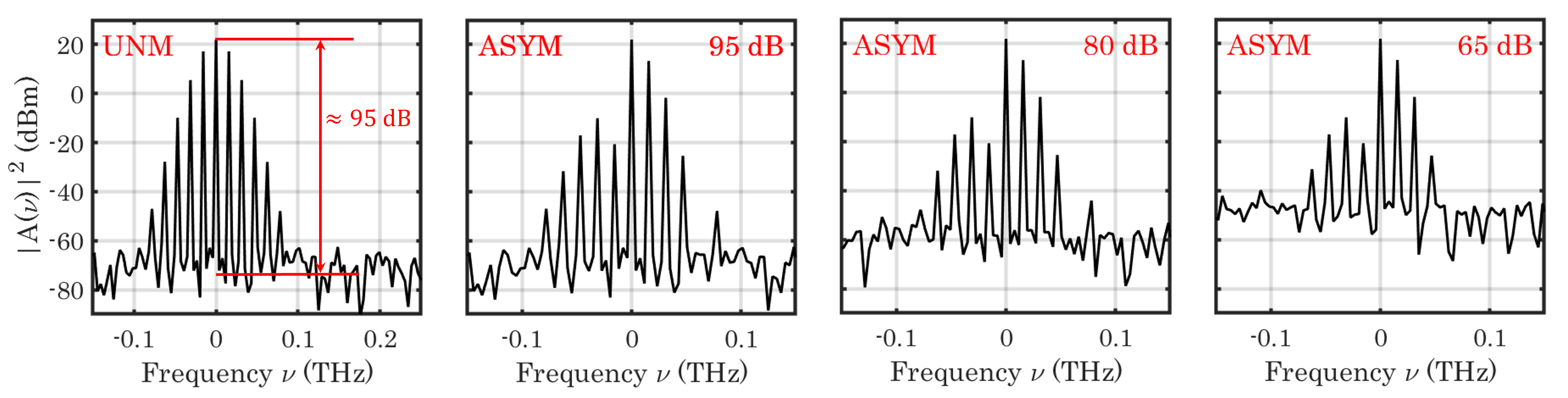}}
\caption{\textcolor{black}{Depiction of the definition of the signal-to-noise ratio (SNR) as a difference between the spectral power of the central comb line and the noise floor as well as examples of a ASYM-encoded frequency comb with three levels of initial noise $(\text{SNR} = 95~\text{dB},$ $80~\text{dB},$ and $65~\text{dB})$ (Eq.~\ref{equ:IC}). The input power is $P_{0}=0.27~\text{W}$ and PM modulation depth is $m = 1.$}}    
\label{fig:DIFF_noise_IC}
\end{figure}

\begin{figure}[bht]
\centering
\fbox{\includegraphics[width= 0.9\textwidth]{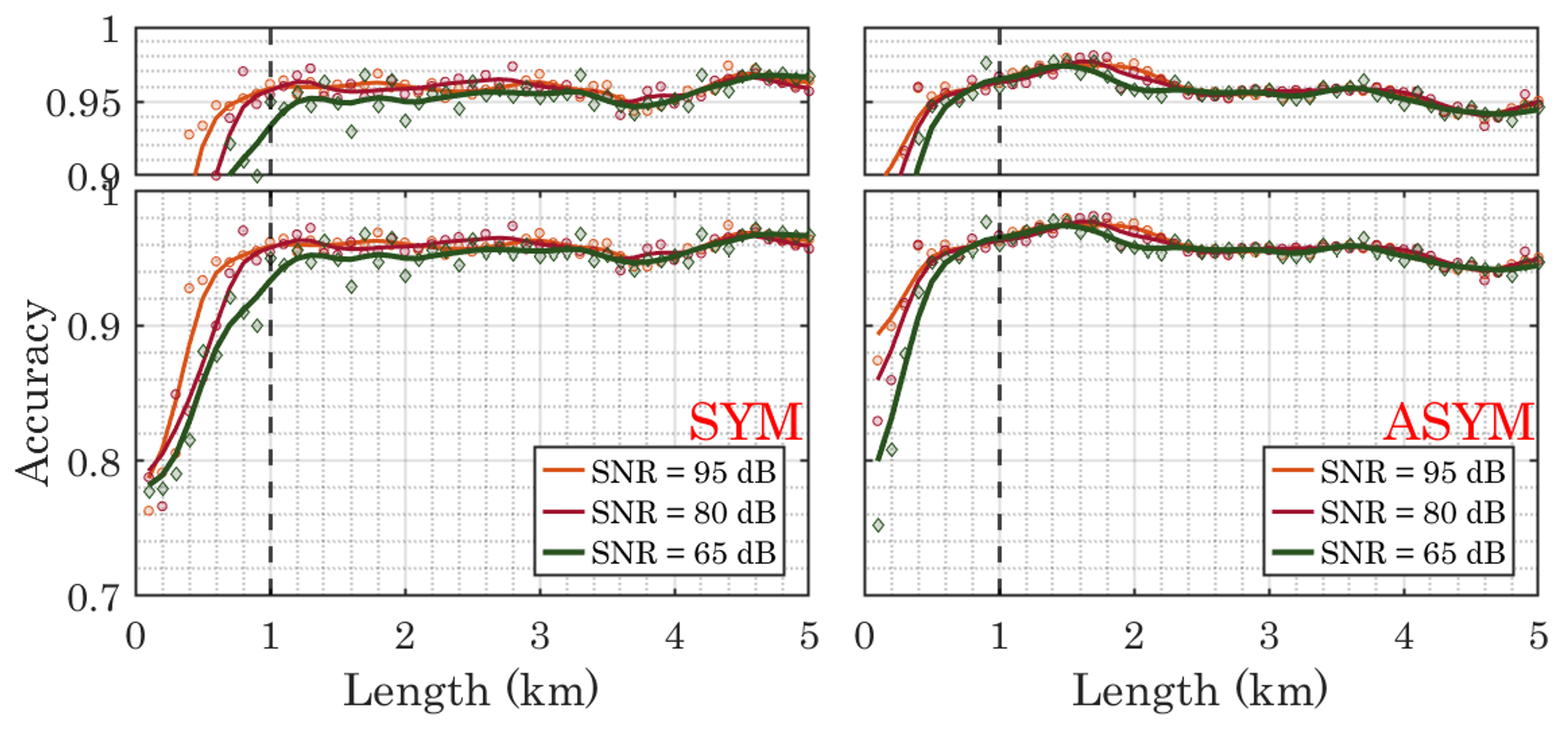}}
\caption{Classification accuracy for Iris dataset over the fiber length of $L = 5~\text{km}$ for symmetric (SYM) and asymmetric (ASYM) information encoding, input power $P_{0}=0.27~\text{W},$ GVD parameter of $\beta_{2} = -23~\text{ps}^{2}/\text{km},$ PM modulation depth $m = 1,$ and different values of initial noise. The averaged standard deviation of the accuracy is $\approx\pm 0.028$ for SYM and for ASYM.}    
\label{fig:DIFF_noise}
\end{figure}

Let us now consider the impact of the initial white noise level on the accuracy performance of the ELM (Eq.~\ref{equ:IC}). This substudy aims to understand how sensitive the ELM is to input noise delivered by the CW laser and the phase modulator PM (Fig.~\ref{fig:SETUP_ENCODING}, top). As noise is usually not a topic of parameter optimization, I will not consider the Shannon entropies of power or phase, but just the accuracy evolution. The assumption to prove is that an increasing level of initial noise would yield degradation of the ELM performance. \textcolor{black}{This assumption is justified, as a high noise floor might "bury" the comb lines, leading to a loss of the encoded information (Fig.~\ref{fig:DIFF_noise_IC}), specifically at low input powers. Below, I consider initial noise levels (Eq.~\ref{equ:IC}) that correspond to spectral signal-to-noise ratio (SNR) of  $\text{SNR} = 95~\text{dB},$ $80~\text{dB},$ and $65~\text{dB}$. From the realization point of view, the first value of $\text{SNR}$ represents a low-noise optical system that would probably require a stabilized laser followed by a low-noise phase modulator PM. The second $\text{SNR}$ value corresponds to a standard optical system that involves a high-quality laser, not necessarily stabilized, and a low-noise phase modulator. The third value of $\text{SNR}$ represents a quite noisy optical system with a more cost-effective laser and phase modulator involved.}

\textcolor{black}{Fig.~\ref{fig:DIFF_noise} shows the evolution of classification accuracy for the SYM and ASYM encoding, input power $P_{0}= 0.27~\text{W},$ and GVD parameter $\beta_{2}= -23~\text{ps}^{2}/\text{km}$ for different initial noise levels. For both types of information encoding, we see a slight degradation of the ELM performance for fiber lengths $L<1~\text{km},$ i.e. the regimes of a steep increase of the accuracy performance over the fiber length. Although this effect is more pronounced for SYM than for ASYM, we are still talking about a good ELM performance, as the accuracy achieves a value of 0.93 at $L=1~\text{km}$ for $\text{SNR} = 65~\text{dB}.$} For ASYM and the same fiber length, there is no difference between the accuracy levels for different initial noise levels. For longer fiber lengths, $L>1~\text{km},$ the effect of the initial noise level is negligible for both, SYM and ASYM, as the accuracy values for one specific initial noise value lie within the variation of accuracy for two other noise values. 

This counterintuitive result is interesting from the point of view of the ELM experimental realization. It indicates that a frequency-multiplexed ELM is robust against initial \textcolor{black}{noise}. As low-noise and stable optical components are usually expensive, ELM robustness to initial noise allows for cost optimization by choosing less expensive, but more noisy and less stable components such as the CW laser and the phase modulator PM. 

\subsection{Breast Cancer Wisconsin Dataset: Shannon-Entropy-Based Optimization in Action}
\begin{figure}[bht]
\centering
\fbox{\includegraphics[width= 0.95\textwidth]{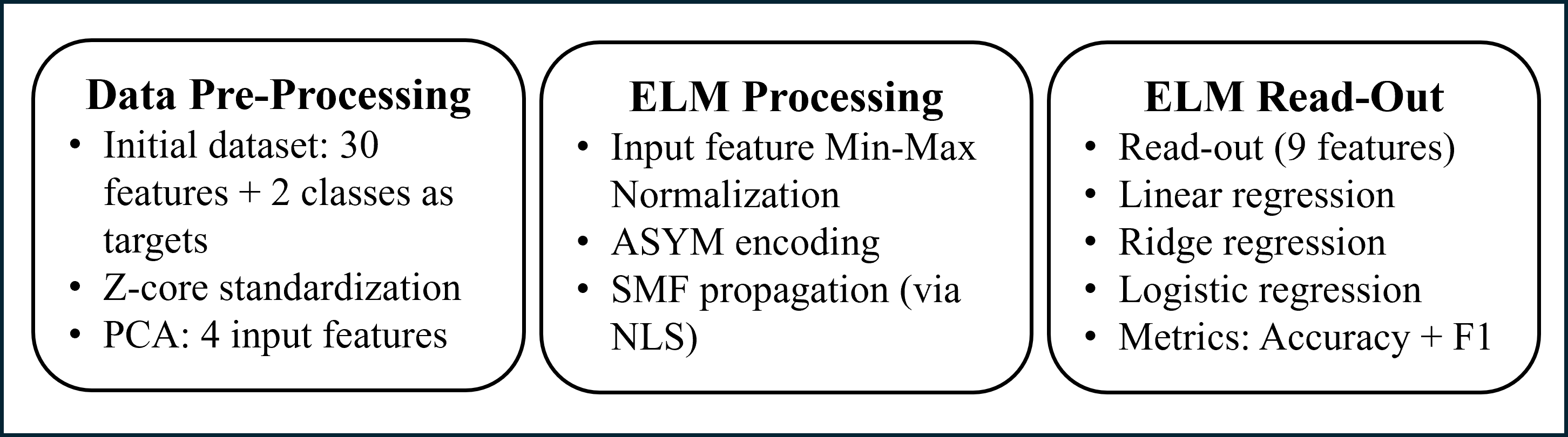}}
\caption{Workflow of processing the Breast Cancer Wisconsin dataset \cite{Wisconsin1993} including data pre-processing and transformation (\textit{left}), data encoding by asymmetric (ASYM) encoding and processing in the SMF stage of the Extreme Learning Machine (Fig.~\ref{fig:SETUP_ENCODING}, top) (\textit{middle}), and read-out with subsequent linear, rigde, and logistic regression (\textit{right}).}    
\label{fig:WORKFLOW_WISCONSIN}
\end{figure}

After having introduced and studied Shannon entropies for optimization of a frequency-multiplexed ELM using the Iris dataset as an example, I now would like to show the effectiveness of this method by comparing an optimized ELM with an unoptimized one using the Breast Cancer Wisconsin dataset \cite{Wisconsin1993}. This dataset belongs to a binary classification problem. Its difficulty consists in being unbalanced with 212 samples of malignant and 357 samples of benign tumor (Sec.~\ref{sec:encoding}). Also, the number of features is 30 which would exceed the number of comb lines for the proposed information encoding (Fig.~\ref{fig:SETUP_ENCODING}, \textcolor{black}{middle}). 

To adjust the number of the input features to the specifications of the ELM, I first standardize the features using the Z-score (also known as standard score) standardization:
\begin{equation}
    \pmb{x}^{j} = \frac{\pmb{x}^{j}_{\text{dataset}}-\mu}{\sigma}
\end{equation}
with $\mu$ being the mean and $\sigma$ the standard deviation of the feature dataset. This standardization is performed to improve the results of the subsequent PCA and to choose 4 principal components that capture $79.24\%$ of the total variance of the dataset. These 4 principal components undergo Min-Max normalization (Eq.~\ref{equ:min_max}). The corresponding values are then ASYM encoded in the frequency comb lines (Sec.~\ref{sec:encoding}) and transformed by propagation through the SMF stage of the ELM. 9 values of the transformed comb line amplitudes are read out and used as features for linear, ridge, and logistic regression to calculate the classification accuracy and the F1 score by evaluating the metrics with a 100-fold cross-validation. A summary of the workflow of the processing of the Breast Cancer Wisconsin dataset is depicted in Fig.~\ref{fig:WORKFLOW_WISCONSIN}.

\begin{figure}[bht]
\centering
\fbox{\includegraphics[width= 1\textwidth]{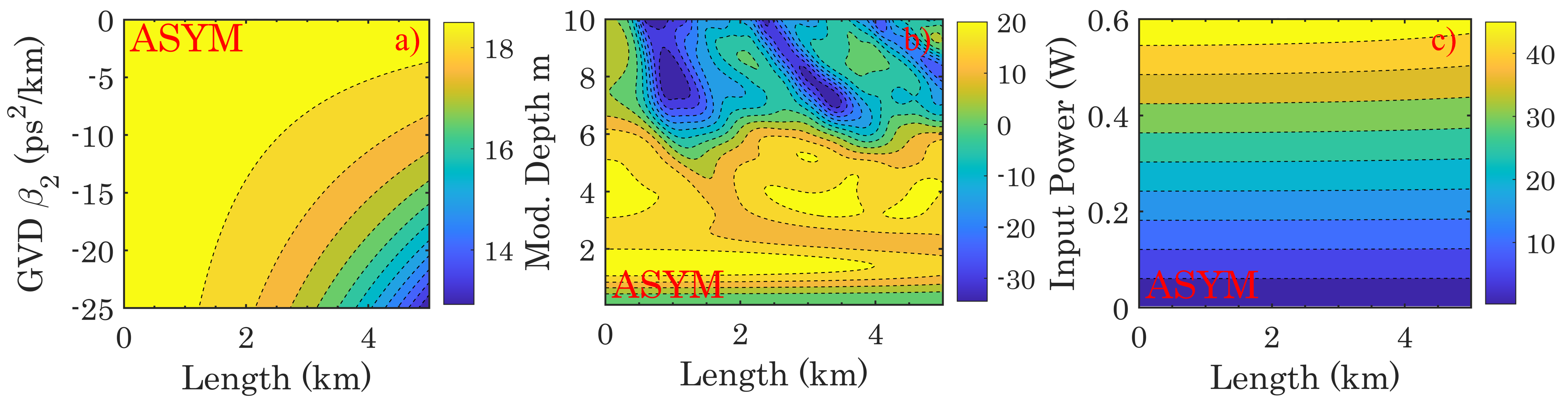}}
\caption{Example of a step-by-step ELM optimization process with Shannon entropy of the optical power. \textit{a)} Scanning for a suitable value of the GVD parameter $\beta_2$ with a fixed optical power of $P_{0}=0.27~\text{W}$ and PM modulation depth $m= 1;$ \textit{b)} scanning for a suitable value of PM modulation depth $m$ with $P_{0}= 0.27~\text{W}$ and GVD parameter $\beta_{2}= -23~\text{ps}^{2}/\text{km};$ \textit{c)} Scanning for input power $P_{0}$ with chosen parameters from a) and b), $\beta_{2}= -5~\text{ps}^{2}/\text{km}$ and $m = 1.2.$ The SMF length is $L = 5~\text{km},$ the power entropy is averaged over $G = 6$ arbitrary samples of the Breast Cancer Wisconsin dataset.
}    \label{fig:Entropy_Wisconsin}
\end{figure}

Fig.~\ref{fig:Entropy_Wisconsin} shows a step-by-step ELM optimization process for the Breast Cancer Wisconsin dataset using Shannon entropy of optical power (Eq.~\ref{eq:Shannon_power}). I use $G=6$ arbitrary values from this dataset (3 values per class) to calculate the entropy.

The optimization process starts with a fixed input power of $P_{0}= 0.27~\text{W}$ and PM modulation depth to scan a suitable value of the GVD parameter $\beta_{2}$. According to Fig.~\ref{fig:Entropy_Wisconsin} (left), a low-dispersion fiber is preferable for an effective ELM. In the next step, I scan the PM modulation depth using $P_{0}= 0.27~\text{W}$ and $\beta_{2}= -23~\text{ps}^{2}/\text{km}.$ According to Fig.~\ref{fig:Entropy_Wisconsin} (middle), values between $m=1.2$ and $m=2$ are preferable. Now, having chosen optimized values $\beta_{2}= -5~\text{ps}^{2}/\text{km}$ and $m=1.2,$ I scan the optical power $P_{0}$ with the result that the best ELM performance \textcolor{black}{is} to expect in the region $P_{0} = 0.55 - 0.60~\text{W}$ (Fig.~\ref{fig:Entropy_Wisconsin}, right). For energy consumption optimization, I choose $P_{0} = 0.55~\text{W}$ as an optimal value. 

\begin{figure}[bht]
\centering
\fbox{\includegraphics[width= 1\textwidth]{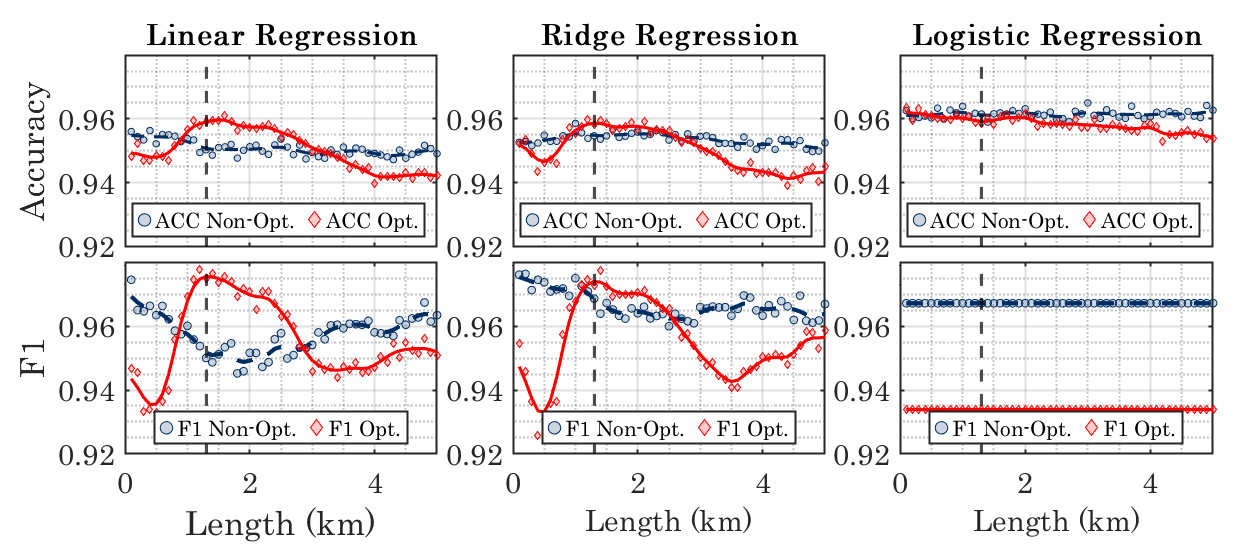}}
\caption{Classification accuracy (\textit{top}) and F1 score (\textit{bottom}) for Linear Regression, Ridge Regression, and Logistic Regression applied at the output layer of the ELM when processing the Breast Cancer Wisconsin dataset over the fiber length of $L = 5~\text{km}.$ The non-optimized, randomly chosen parameter set (blue curves) consists of $\beta_{2} = -25~\text{ps}^{2}/\text{km},$ $m = 3$ and $P_{0}= 0.2~\text{W};$ the optimized one (red curves) of $\beta_{2} = -5~\text{ps}^{2}/\text{km},$ $m = 1.2$ and $P_{0}= 0.55~\text{W}.$}    
\label{fig:ACC_Wisconsin}
\end{figure}

Now, with optimized values of $\beta_{2}= -5~\text{ps}^{2}/\text{km}$ and $m=1.2,$ and $P_{0} = 0.55~\text{W},$ I calculate the overall classification accuracy and F1 score using linear, ridge, and logistic regression. After several performance optimization runs, the best regularization parameter for the ridge regression is found to be $\alpha = 0.5.$ The concept of ELM includes using linear regression models to optimize the weights of the output layer and calculate the results of the ELM read-out layer (\cite{Zajnulina2023}, \cite{Lupo2021}, \cite{Huang2014}, \cite{Ding2014}). I added the (nonlinear) logistic regression to test whether it can improve the classification accuracy and F1 \textcolor{black}{score.}

The results of an optimized ELM are shown in red in Fig.~\ref{fig:ACC_Wisconsin}. For a comparison, I also show the results in blue that are achieved for an unopmitized ELM with randomly chosen, but experimentally realistic, values of $\beta_{2}= -25~\text{ps}^{2}/\text{km}$ and $m=3,$ and $P_{0} = 0.2~\text{W}.$ As we see, the optimized ELM outperforms the non-optimized one for the case of linear and ridge regression with accuracy values of $0.9592$ (linear regression) and $0.9582$ (ridge regression) at SMF length of $L = 1.3~\text{km}.$ The corresponding F1 score values are $0.975$ and $0.973.$ Further, we see that the ELM optimization via Shannon entropy of optical power not only improves the performance, but also gives us some freedom in the choice of the SMF length simplifying the experimental realization. Thus, best values of the optimized ELM can be found for SMF length of $L = 1.0 - 2.5~\text{km}$ with a maximum at $L = 1.3~\text{km}.$  

Ref.~\cite{Naji_2021} compares different machine learning models for the Breast Cancer Wisconsin dataset and reports a classification testing-dataset accuracy of 0.937 as the lowest value (K-nearest-neighbors) and 0.972 as the highest value (Support Vector Machine). Accuracy values of the Shannon-entropy optimized ELM achieved here, i.e. $0.9592$ (linear regression) and $0.9582$ (ridge regression), place it among the top-performing state-of-the-art classification models for this dataset although only 4 principal components were used to encode the information and produce the classification output of the ELM. A higher number of principal components is expected to deliver even better results. 

Interestingly, when logistic regression is applied at the output layer of the ELM, the non-optimized version performs better than the optimized one with an accuracy value of $0.9618$ and F1 score of $0.9672$ at $L= 1.3~\text{km}.$ These results top the logistic regression accuracy of $0.958$ (on test dataset) presented in Ref.~\cite{Naji_2021}. An ELM equipped with a logistic regression at the output layer can be seen as a pipeline of two consecutive nonlinear machine learning models which might explain the improved result of the logistic-regression ELM as compared just to a logistic regression model (\cite{Naji_2021}). On the other hand, an optimized ELM in connection with a logistic regression model at the output, can be seen as over-trained which would explain the decrease in the accuracy and F1 values for the optimized ELM. Certainly, further studies are needed to better understand the applicability and feasibility of ELM output-layer models other than the linear ones (linear and ridge regression).   

\section{Conclusion}
\label{sec:conclusion}
I consider an Extreme Learning Machine (ELM) that utilizes frequency multiplexing to encode data in the line amplitudes of a frequency comb and \textcolor{black}{processes} these data in a standard single-mode fiber subject to Kerr nonlinearity. To derive conclusions about the internal dynamics and exploit these conclusions to optimize the ELM performance, I introduce the notions of Shannon entropy of optical power, phase, and spectrum and, using numerical simulations of nonlinear light propagation in the ELM, show its effectiveness as an optimization tool. 

The following system parameters were considered: continuous-laser optical power, fiber group-velocity dispersion, the modulation depth of the phase modulator that produces an initial frequency comb out \textcolor{black}{of} the laser radiation, and two types of information encoding, symmetric and asymmetric in frequency-comb lines. Two datasets were used to evaluate the performance of the ELM: Iris and Breast Cancer Wisconsin datasets. A comparison between Shannon entropy and classification accuracy plots revealed that the highest entropy values relate to the fast increases of the accuracy values to their maxima while decreasing entropy coincides with the stagnation and decrease of the accuracy. As a result, the best ELM performance can be expected for continuous-wave laser optical powers of $0.45 - 0.6~\text{W},$ phase-modulator modulation depth $m<3,$ and low-dispersive fibers with fiber lengths of $0.8 - 1.3~\text{km}.$ As low modulation depth generates a number of frequency comb lines that could be too little to encode datasets with a large number of features, Principal Component Anasysis can be used to reduce the dimensionality of the feature space. With this approach, I reduce the dimension of the Breast Cancer Wisconsin dataset from 30 features per sample to 4 and show that an ELM optimized by Shanon entropy of optical power and phase yield classification results that place the ELM among the top-performing state-of-the-art machine learning schemes for this dataset. Apart from that, it was shown that the ELM is robust with respect to the initial noise. These results pave the way for cost-effective, simplified system designs (in fibers and on-chip) operating at potentially lower energy costs than the ones that are available now (cf.~\cite{Zajnulina2023}). 

Two schemes of initial information encoding, symmetric and asymmetric in frequency-comb lines, are introduced and their impact on ELM dynamics and performance is studied. Coinciding with stronger Four Wave Mixing (FWM), asymmetric information encoding performs slightly better than symmetric one which supports the theory of FWM being the main mechanism for information processing in a frequency-multiplexed ELM (\cite{Zajnulina2023}). Symmetric encoding rather supports soliton-driven information processing (\cite{Marcucci2020}). However, FWM and the formation of solitonic waves are not separable effects as they take place in parallel, their isolated impact on ELM performance remains unrevealed and, with the availability of Shannon entropy as a tool, not essential for optmization and control of the ELM. 

To gain more understanding of what nonlinear dynamics might be generated in the single-mode fiber by a frequency comb modulated with symmetric and asymmetric encoding, other methodology needs to be applied. To do so, I deployed Soliton Radiation Beat Analysis, a numerical technique that allows retrieving soliton content from arbitrary inputs \cite{Boehm_2006}. I used the first sample of the Iris dataset to modulate the initial frequency comb produced by phase modulation of the continuous-wave laser radiation. Thus, symmetric modulation leads to an input-power-dependent evolution of optical structures similar to Akhmediev breathers and separated Peregrine solitons. Asymmetric modulation of the frequency comb yields the formation of soliton crystals additionally to Akhmediev-breathers-like structures and separated solitons. For very low input powers, symmetric and asymmetric frequency combs are subject to the quasi-linear temporal Talbot effect \cite{Zajnulina2024}. These findings contribute to a better understanding of possible dynamics of modulated frequency combs in optical fibers and constitute, to the best of my knowledge, the first reported attempt to retrieve the solitonic-wave \textcolor{black}{type and content} from modulated frequency combs in such detail. Therefore, it is relevant to the field of Nonlinear Optics. Certainly, further studies are needed to gain deeper insights into the behavior of modulated frequency combs in optical fibers and Kerr media in general. 

To conclude, the introduced Shannon entropy of optical power, phase, and spectrum is a promising tool for effective Extreme Learning Machine \textcolor{black}{design and performance} optimization without the need to precisely \textcolor{black}{know and} understand the internal dynamics of the ELM as \textcolor{black}{the entropies} serve as a dynamical indicator \textcolor{black}{themselves. Providing (indirect) insights into the ELM's internal dynamics, this method enhances the development of approaches for explainable AI (XAI) in optical computing. It requires comparably low computational resources and time and shows a high potential of applicability to other guided ELM schemes in optical fibers and semiconductor substrates.} For uncovering the dynamics \textcolor{black}{in a more targeted and informative way}, Soliton Radiation Beat Analysis can be applied. Together, \textcolor{black}{these methods} enrich the toolset of Neuromorphic Photonics and Nonlinear Optics in general.

\section*{Funding}
Project Win4Space / Win4ReLaunch (SPW EER Wallonie Belgique, grant agreement number 2210181).

\section*{Acknowledgments} I acknowledge inspiring conversations about Extreme Learning Machines with Serge Massar (Université libre de Bruxelles) and would like to say thank you to Michael B\"ohm (Wismar University of Applied Sciences), the inventor of Soliton Radiation Beat Analysis, for our long and highly interesting conversations about optical solitons.

\section*{Disclosures} I declare no conflict of interest.

\section*{Data availability} Data underlying the results presented in this paper are not publicly available at this time but may be obtained from me upon reasonable request.

\appendix
\section{Nonlinear Evolution of Optical Combs Modulated via SYM and ASYM Encoding}
\label{sec:app1}

\begin{figure}[th]
\centering
\fbox{\includegraphics[width= 0.98\textwidth]{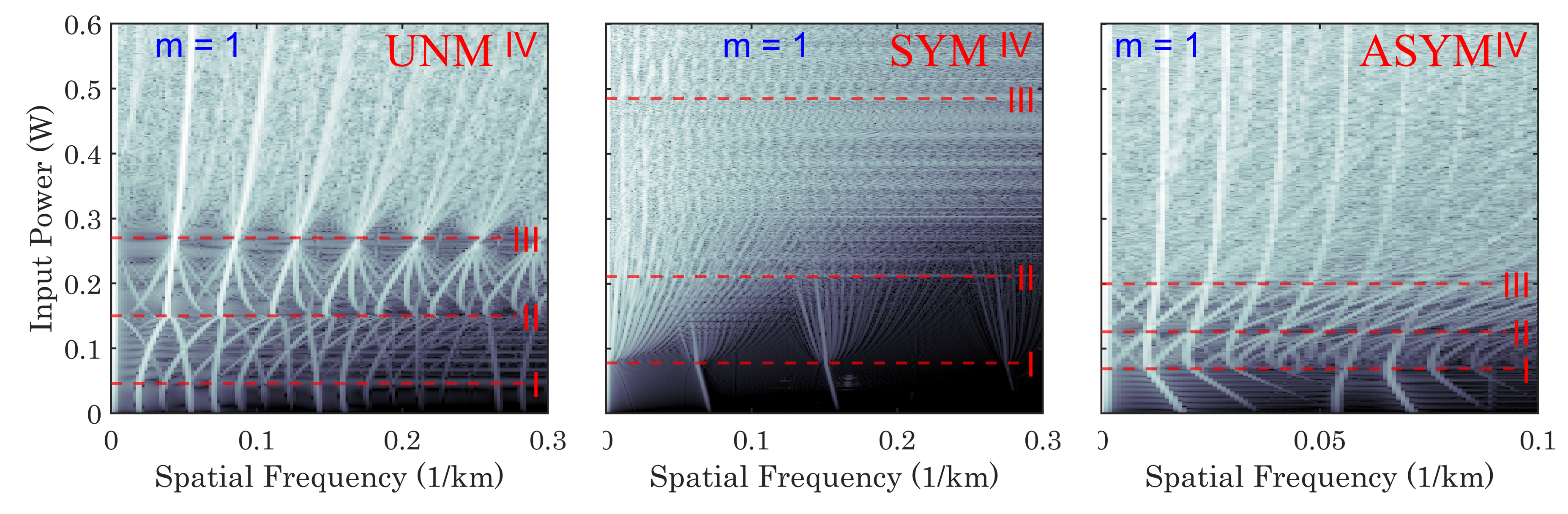}}
\caption{Soliton Radiation Beat Analysis spatial power spectrum  in \textcolor{black}{dB} obtained for an SMF with GVD parameter $\beta_2 = -23~\text{ps}^{2}/\text{km}$ and nonlinear coefficient $\gamma = 1.2~(\text{W}\cdot\text{km})^{-1}$ \textit{Left:} for an unmodulated (UNM) initial frequency comb (\cite{Zajnulina2024}), \textit{center:} a comb modulated via SYM, and \textit{right:} a comb modulated via ASYM with the first sample from Iris dataset.}
\label{fig:SRBA_OPT}
\end{figure} 

\begin{figure}[th]
\centering
\fbox{\includegraphics[width= 0.90\textwidth]{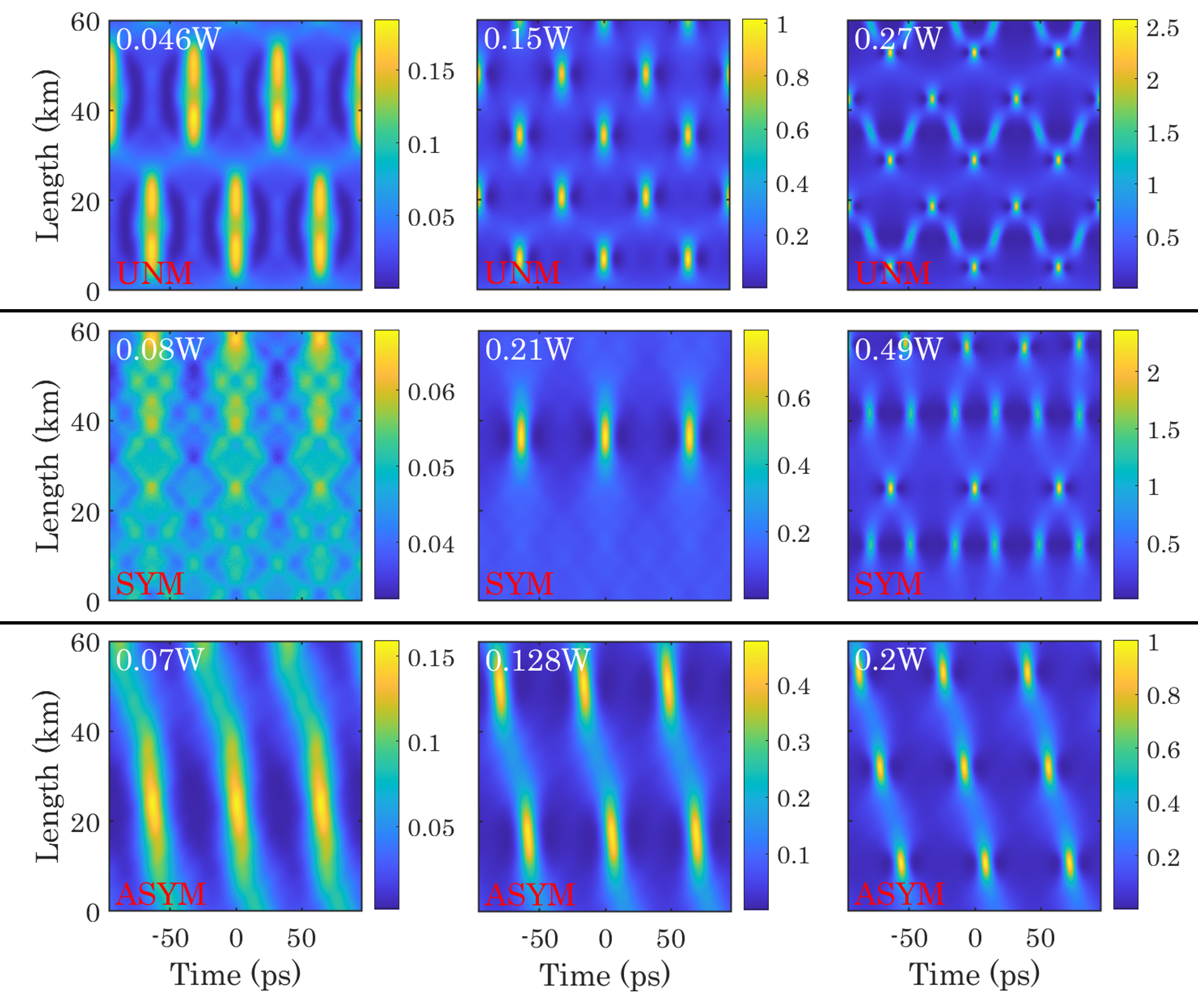}}
\caption{\textcolor{black}{Optical power evolution for an SMF with GVD parameter $\beta_2 = -23~\text{ps}^{2}/\text{km}$ and nonlinear coefficient $\gamma = 1.2~(\text{W}\cdot\text{km})^{-1}$ for an unmodulated (UNM) initial frequency comb \textit{(top)} (\cite{Zajnulina2024}), a comb modulated via SYM \textit{(middle)}, and a comb modulated via ASYM \textit{(bottom)} with the first sample from Iris dataset.}}
\label{fig:OPT_DYNAMCIS}
\end{figure} 

This section provides insight into possible dynamics that evolve when a comb modulated by symmetric (SYM) or asymmetric (ASYM) encoding is injected into a piece of SMF (Sect.~\ref{sec:encoding}). It aims at readers who are particularly interested in nonlinear dynamics. Thorough knowledge and understanding of each detail are not needed to grasp the working principle of the ELM (Fig.~\ref{fig:SETUP_ENCODING}, top). Apart from that, the dynamics of such systems is an ongoing research and many open questions remain unanswered. Here, I attempt to decode possible dynamics using the so-called \textit{Soliton Radiation Beat Analysis} (SRBA) (\cite{Boehm_2006}, \cite{Zajnulina_2015}, \cite{Zajnulina_2017}, \cite{Mitschke2017}). It will allow us to see different input-power dependent regimes taking place in an SFM for an unmodulated (UNM) frequency comb as an input as well as a comb modulated symmetrically (SYM) and asymmetrically (ASYM). It is the first glance of this kind, and more profound follow-up studies will be needed to better understand the regimes.

The SRBA is a numerical technique to retrieve the soliton content of pulses generated in optical fibers from arbitrary inputs. It is done by quantitative analysis of spatial frequencies of optical-power oscillations that arise due to higher-order soliton oscillations, or beating between several co-existent solitons, or beating between solitons and dispersive waves \cite{Boehm_2006}, \cite{Zajnulina_2015}, \cite{Zajnulina_2017}, \cite{Mitschke2017}. 

The SRBA includes the following steps: i) numerical integration of the NLS (Eqs.~\ref{equ:NLS}, \ref{equ:IC}) to calculate the optical field \textcolor{black}{$A(z,t)$} and the optical power \textcolor{black}{$|A(z,t)|^{2}$} \textcolor{black}{in W}; ii) extraction of the optical power at a certain time point \textcolor{black}{$t_{0},$} for instance, \textcolor{black}{$t_{0} = 0~\text{ps}$} being the center of the chosen time window; and iii) fast Fourier transform of the optical power at this time point and the calculation of the power spectrum of spatial frequencies in \textcolor{black}{dB}. Calculated for different values of the input power $P_{0},$ these spatial-frequency power spectra are stacked into a 2D (spatial frequency - input power) SRBA plot. 

To interpret SRBA plots, one needs to bear in mind the phase evolution of different waves in optical fibers. Thus, the phase of a linear (for instance, dispersive) wave evolves mainly depending on GVD and modulation frequency $\Omega:$ $\phi_{lin}(z) \propto\frac{\beta_2\Omega^{2}}{2}z.$ The phase of solitons primarily depends on the input power $P:$ $\phi_{nl}(z) \propto \gamma P z.$ Accordingly, linear waves are recognizable as power-independent \textit{vertical} lines with fixed spatial frequencies, whereas soliton spatial frequencies constitute input-power dependent rather \textit{parabolic} branches in the SRBA plot \cite{Boehm_2006} \textcolor{black}{(cf. Fig.~\ref{fig:SRBA_OPT}).} This dependence implies that the oscillation period of solitonic waves will nonlinearly decrease with \textcolor{black}{increasing} input power. 

The resolution of SRBA plots depends on the fiber propagation length: the longer the fiber, the sharper and better visible the spatial-frequencies structures. Thus, here, I use an SMF length of $L = 500~\text{km}$ for the integration of Eqs.~\ref{equ:NLS}, \ref{equ:IC}. Other parameters are: $\beta_2 = -23~\text{ps}^{2}/\text{km}$ and $\gamma = 1.2~(\text{W}\cdot\text{km})^{-1}.$ Higher-order dispersion, optical losses, shock, and the Raman effect are excluded for simplicity's sake. The first example of the Iris dataset is used to modulate the initial comb via SYM and ASYM encoding. As the variance of the Iris dataset samples is quite small, SRBA plots for other samples are expected to look similar with small differences in the threshold input-power values that separate different dynamical regimes.

Ref.~\cite{Zajnulina2024} discusses different power-dependent quasi-linear and nonlinear regimes in an SMF revealed by the SRBA of an unmodulated \textcolor{black}{comb} (Figs.~\ref{fig:SRBA_OPT} left and \ref{fig:OPT_DYNAMCIS} top). Thus, for low input powers $(P_{0} = 0.0 - 0.046~\text{W})$ (regime I), the dynamics is governed by quasi-linear temporal Talbot effect (\cite{Jannson_1981}, \cite{Wu_2023}): the spatial frequencies are almost independent of the value of $P_{0}.$ In regime II $(P_{0} = 0.046 - 0.15~\text{W}),$ the spatial frequencies acquire nonlinear-phase contributions growing together under momentum conservation and building Akhmediev-breather-like structures (cf. \cite{Kimmoun_2016}) manifesting themselves in pitchfork-like spatial frequencies. These structures are transformed into soliton crystals in regime III $(P_{0} = 0.15 - 0.27~\text{W})$ with fan-like spatial frequencies as a characteristic (\cite{Zajnulina_2015}, \cite{Zajnulina_2017}). In regime IV $(P_{0} \geq 0.27~\text{W}),$ the optical pulses form separated solitons recognizable by parabolic branches in the SRBA plot. Akhmediev-breather-like structures and soliton crystals constitute spatio-temporal compounds, whereas separated solitons are independent and hardly interact with each other due to their spatio-temporal localization. 

With SYM encoding deployed, the dynamics look differently (Figs.~\ref{fig:SRBA_OPT} middle and \ref{fig:OPT_DYNAMCIS} middle). In regime I $(P_{0} = 0.0 - 0.08~\text{W}),$ the dynamics is still governed by the quasi-linear Talbot effect. A wider span of the input power for this regime (as compared to the UNM case) results from the decrease of the average power due to information encoding (cf. Fig.~\ref{fig:SETUP_ENCODING}, bottom). At $P_{0} = 0.08~\text{W},$ we observe a transition to regime II, where structures arise that can be interpreted as Peregrine solitons who constitute spatiotemporally localized limit cases of Akhmediev breathers (\cite{Hammani_2011}, \cite{Frisquet2013}, \cite{Ye_2020}). We can see a train of such Peregrine solitons at ca. $L=39~\text{km},$ $P_{0} = 0.21~\text{W}$ in Fig.~\ref{fig:OPT_DYNAMCIS}, middle. In regime III $(P_{0} = 0.21 - 0.49~\text{W}),$ Peregrine-soliton structures (re-)appear at propagation distances that decrease with input power, giving rise to a dense pattern of parabolic branches in the SRBA plot (Fig.~\ref{fig:SRBA_OPT}, middle). For $P_{0} \geq 0.49~\text{W}),$ we see that these dense branches split into sub-brunches, which can be interpreted as Peregrine solitons emitting separated solitons. It is worth pointing out that the regime of soliton crystals is missing for SYM.

For ASYM (Figs.~\ref{fig:SRBA_OPT}, right, and \ref{fig:OPT_DYNAMCIS}, bottom), the regime I governed mainly by the quasi-linear temporal Talbot effect takes place for  $P_{0} = 0.0 - 0.07~\text{W}.$ In regime II $(P_{0} = 0.07 - 0.128~\text{W}),$ we see some SRBA pitchfork structures that can be associated with the formation of Akhmediev-breather-like waves (\cite{Zajnulina2024}). Also, we see power-dependent branches that start at spatial frequency $0~\text{km}^{-1}$ and dissolve from the pitchforks. Those are markers for separated solitons (\cite{Boehm_2006}). It means that the optical pulses constitute a beating of Akhmediev-breather-like waves with separated solitons. In regime III $(P_{0} = 0.128 - 0.2~\text{W}),$ the beating of separated solitons (power-dependent branches thresholding, among others, at $0~\text{km}^{-1}$) occurs with soliton crystals recognizable as fans of spatial frequencies. In regime IV $P_{0}\geq0.2~\text{W},$ we see branches of separated solitons with their typical parabolic shapes (\cite{Boehm_2006}, \cite{Zajnulina_2015}, \cite{Zajnulina_2017}, \cite{Mitschke2017}). Interestingly, the regimes of Akhmediev-like structures (II) and soliton crystals (III) cover a smaller input-power range and dissolve into separated solitons at lower input power $(P_{0}=0.2~\text{W})$ than in the UNM case. A possible explanation can relate to the asymmetry of the comb: as the ASYM-modulated comb propagates through the SMF, it tries to balance its spectral energy distribution by compensating the asymmetry via enhanced FWM. Enhanced FWM, in turn, helps create more temporally compressed pulses, which eventually leads to the formation of temporally localized separated solitons. More studies are certainly needed to better understand this process. In all regimes, the initial asymmetry of the frequency comb causes a temporal shift of the pulses \cite{Schiek2021}.

As we have seen in this section, as soon as the input delivers enough optical power to transit from quasi-linear (\cite{Jannson_1981}) to nonlinear temporal Talbot effect (\cite{Cohen_2008}, \cite{Wen_2011}, \cite{Zhang_2014},) we observe the development of some sort of solitonic (Akhmediev-breather-like, crystals, or separated solitons) optical structures in the SMF, with ASYM inputs being more prone to FWM than the SYM inputs (Sec.~\ref{sec:Dynamics_Note}). This development takes place over a range of several kilometers. For the ELM, I consider SMF lengths of $L\leq 5~\text{km}$ only. As these lengths are not sufficient to build full-scale solitonic waves for most considered input powers (apart from powers $P_{0}\rightarrow0.6~\text{W}$ for ASYM), I would rather call these waves proto-solitonic, as long as they achieve the input-power value of the transition from quasi-linear to nonlinear temporal Talbot effect. The input powers of the nonlinear Talbot effect correlate with rapid increase and, in general, higher accuracy values of the ELM (Sec.~\ref{sec:results}). 

Concerning the ELM's working principle, it is important to note that it is not possible to separate the proto-solitonic wave development from FWM as a mechanism for data processing as these effects occur simultaneously (in particular in the case of ASYM). In these terms, as long as there is (anomalous) dispersion present in the fiber (that facilitates the formation of proto-solitonic waves), both processes, soliton/breather/crystal formation and FWM, are involved in the ELM data-processing capabilities (\cite{Zajnulina2023}, \cite{Marcucci2020}).


\begin{thebibliography}{00}


\bibitem{McMahon2023}
    P. L. McMahon,
    \textit{The physics of optical computing},
    Nat. Rev. Phys. Vol. 5(12), pp. 717-734, 2023, 
    \url{https://doi.org/10.1038/s42254-023-00645-5}.

\bibitem{Shastri2021}
    B.J. Shastri, A.N. Tait, T. Ferreira de Lima, W. H. P. Pernice, H. Bhaskaran, C. D. Wright, P. R. Prucnal, 
    \textit{Photonics for artificial intelligence and neuromorphic computing}, 
    Nat. Photonics Vol. 15, pp. 102–-114, 2021, \url{https://doi.org/10.1038/s41566-020-00754-y}.

\bibitem{Pauwels2019}
    J. Pauwels, G. Verschaffelt, S. Massar, G. Van der Sande
    \textit{Distributed Kerr Non-linearity in a Coherent All-Optical Fiber-Ring Reservoir Computer}
    Front. Phys. Vol. 7, 2019, \url{https://www.frontiersin.org/journals/physics/articles/10.3389/fphy.2019.00138}.

\bibitem{Fischer2023}
    B. Fischer, M. Chemnitz, Y. Zhu, N. Perron, P. Roztocki, B. MacLellan, L. Di Lauro, Luigi, A Aadhi, C. Rimoldi, T. H. Falk, R. Morandotti,
    \textit{Neuromorphic Computing via Fission-based Broadband Frequency Generation},
    Adv. Sci. Vol. 10(35), 2303835, 2023,
    \url{https://doi.org/10.1002/advs.202303835}.

\bibitem{Zajnulina2023}
         M. Zajnulina, A. Lupo, S. Massar,
         \textit{Weak Kerr nonlinearity boosts the performance of frequency-multiplexed photonic extreme learning machines: a multifaceted approach.}
         Opt. Express Vol. 33(4), pp. 7601-7619, 2025,
         \url{https://doi.org/10.1364/OE.503279}.

\bibitem{Lupo2021}
    A. Lupo, L. Butschek, S. Massar,
    \textit{Photonic extreme learning machine based on frequency multiplexing.}
    Opt. Express  Vol. 29(18), pp. 28257-28276, 2021,
    \url{https://doi.org/10.1364/OE.433535}.

\bibitem{Brunner2018}
    D. Brunner, B. Penkovsky, B. A. Marquez, M. Jacquot, I. Fischer, L. Larger,
    \textit{Tutorial: Photonic neural networks in delay systems},
    J. Appl. Phys. 124, 152004, 2018,
    \url{https://doi.org/10.1063/1.5042342}.    

\bibitem{Phang2023}
    S. Phang, 
    \textit{Photonic reservoir computing enabled by stimulated Brillouin scattering,}
    Opt. Express Vol. 31(13), pp. 22061--22074, 2023,
    \url{https://doi.org/10.1364/OE.489057}.

\bibitem{Oguz2024}
    I. Oguz, J.-L. Hsieh, N. Ulas Dinc, U. Teğin, M. Yildirim, C. Gigli, C. Moser, D. Psaltis,
    \textit{Programming nonlinear propagation for efficient optical learning machines},
    Adv. Photonics Vol. 6(1), 016002, 2024,
    \url{https://doi.org/10.1117/1.AP.6.1.0160020}.

\bibitem{Yildrim2023}
    M. Yildirim, N. U. Dinc, I. Oguz, D. Psaltis, C. Moser,
    \textit{Nonlinear processing with linear optics},
    Nat. Photon. Vol. 18, pp. 1076-–1082, 2024, \url{https://doi.org/10.1038/s41566-024-01494-z}.

\bibitem{Wetzstein2020}
    G. Wetzstein, A. Ozcan, S. Gigan, S. Fan, D. Englund, M. Soljacic, C. Denz, D. A. B. Miller, D. Psaltis,
    \textit{Inference in artificial intelligence with deep optics and photonics},
    Nature 588, pp. 39-–47, 2020, \url{https://doi.org/10.1038/s41586-020-2973-6}.

\bibitem{Redding2024}
    B. Redding, J. B. Murray, J. D. Hart, Z. Zhu, S. S. Pang, R. Sarma,
    \textit{Fiber optic computing using distributed feedback}, Commun Phys 7(75), 2024,
    \url{https://doi.org/10.1038/s42005-024-01549-1}.

\bibitem{Cox2024}
    N. Cox, J. Murray, J. Hart, B. Redding, 
    \textit{Photonic next-generation reservoir computer based on distributed feedback in optical fiber},
    Chaos 34(7), 073111, 2024, 
    \url{https://doi.org/10.1063/5.0212158}.

\bibitem{Wanjura2023}
    C. C. Wanjura, F. Marquardt,
    \textit{Fully nonlinear neuromorphic computing with linear wave scattering,}
    Nat. Phys. Vol. 20, pp. 1434-–1440, 2024, \url{https://doi.org/10.1038/s41567-024-02534-9}.

\bibitem{Xia2024}
    F. Xia, K. Kim, Y. Eliezer, SY. Han, L. Shaughnessy, S. Gigan, H. Cao,
    \textit{Nonlinear optical encoding enabled by recurrent linear scattering,}
    Nat. Photon. 18, pp. 1067-–1075, 2024, \url{https://doi.org/10.1038/s41566-024-01493-0}.

\bibitem{Bai2023}
    Y. Bai, X. Xu, M. Tan, Y. Sun, Y. Li, J. Wu, R. Morandotti, A. Mitchell, K. Xu, D. J. Moss
    \textit{Photonic multiplexing techniques for neuromorphic computing},
    Nanophotonics Vol. 12(5), pp. 795--817, 2023,     
    \url{https://doi.org/10.1515/nanoph-2022-0485}.

\bibitem{Xu2023_ONN}
    X. Xu, W. Han, M. Tan, Y. Sun, Y. Li, J. Wu, R. Morandotti, A. Mitchell, K. Xu, D. J. Moss,
    \textit{Neuromorphic Computing Based on Wavelength-Division Multiplexing},
    IEEE J. Sel. Top. Quantum Electron. Vol. 29(2), 2023, 
    \url{10.1109/JSTQE.2022.3203159}.

\bibitem{Argyris2018}
    A. Argyris, J. Bueno, I. Fischer,
    \textit{Photonic machine learning implementation for signal recovery in optical communications.}
    Sci. Rep. 8, 8487, 2018,
    \url{https://doi.org/10.1038/s41598-018-26927-y}.

\bibitem{Duport2016}
    F. Duport, A. Smerieri, A. Akrout, M. Haelterman, S. Massar,
    \textit{Fully analogue photonic reservoir computer.}
    Sci. Rep. 6, 22381, 2016,
    \url{https://doi.org/10.1038/srep22381}.

\bibitem{Butschek2022}
    L. Butschek, A. Akrout, E. Dimitriadou, A. Lupo, M. Haelterman, S. Massar,
    \textit{Photonic reservoir computer based on frequency multiplexing.}
     Opt. Lett. Vol. 47(4), pp. 782-785, 2022, 
     \url{https://doi.org/10.1364/OL.451087.}

\bibitem{Suzuki2022}
    Y. Suzuki, Q. Gao, K. C. Pradel, K. Yasuoka, N. Yamamoto,
    \textit{Natural quantum reservoir computing for temporal information processing.}
    Sci. Rep. 12, 1353, 2022,
    \url{https://doi.org/10.1038/s41598-022-05061-w}.


\bibitem{Huang2014}
    G.-B. Huang,
    \textit{An Insight into Extreme Learning Machines: Random Neurons, Random Features and Kernels},
    Cogn. Comput. Vol. 6(3), pp. 376-390, 2014,
    \url{https://doi.org/10.1007/s12559-014-9255-2}.


\bibitem{Ding2014}
    S. Ding, X. Xu, R. Nie,
    \textit{Extreme learning machine and its applications.}
    Neural Comput. \& Applic. Vol. 25, pp. 549-556, 2014,
    \url{https://doi.org/10.1007/s00521-013-1522-8}.

\bibitem{Duarte2023}
    D. Silva, T. Ferreira, F. C. Moreira, C. C. Rosa, Carla, A. Guerreiro, N. A. Silva,
    \textit{Exploring the hidden dimensions of an optical extreme learning machine},
    J. Eur. Opt. Society-Rapid Publ. Vol. 19(1), p. 8, 2023, 
    \url{https://doi.org/10.1051/jeos/2023001}.

\bibitem{Zhou2022}
  T. Zhou, Tingyi, F. Scalzo, B. Jalali,
  \textit{Nonlinear Schrödinger Kernel for Hardware Acceleration of Machine Learning}, 
  J. Light. Technol. Vol. 40(5), pp. 1308-1319, 2022,
  \url{10.1109/JLT.2022.3146131}.


\bibitem{Yildirim2023}
    M. Yildirim, I. Oguz, F. Kaufmann, M. R. Escalé, R. Grange, D. Psaltis, C. Moser,
    \textit{Nonlinear optical feature generator for machine learning},
    APL Photonics Vol. 8(10), p. 106104, 2023,
    \url{https://doi.org/10.1063/5.0158611}.


\bibitem{Oguz2023}
    I. Oguz, J. Ke, Q. Weng, F. Yang, M. Yildirim, N. Ulas Dinc, J.-L. Hsieh, C. Moser, D. Psaltis,
    Opt. Lett. Vol. 48(20), pp. 5249--5252, 2023,
    \url{https://opg.optica.org/ol/abstract.cfm?URI=ol-48-20-5249}.

\bibitem{Skontranis2023}
    M. Skontranis, G. Sarantoglou, K. Sozos, T. Kamalakis, C. Mesaritakis, A. Borgis,  
    \textit{Multimode Fabry-Perot laser as a reservoir computing and extreme learning machine photonic accelerator},
    Neuromorph. Comput. Eng. 3, 044003, 2023,
    \url{https://doi.org/10.1088/2634-4386/ad025b}.

\bibitem{Biasi2023}
    S. Biasi, R. Franchi, L. Cerini, L. Pavesi,
    \textit{An array of microresonators as a photonic extreme learning machine.}
    APL Photon. Vol. 8, 096105, 2023,
    \url{https://doi.org/10.1063/5.0156189}.

\bibitem{Pierangeli2021}
    D. Pierangeli, G. Marcucci, C Conti, 
    \textit{Photonic extreme learning machine by free-space optical propagation.} Photon. Res. 9, pp. 1446-1454, 2021,
    \url{https://doi.org/10.1364/PRJ.423531}.

\bibitem{Marcucci2020}
    G. Marcucci, D. Pierangeli, C. Conti,
    \textit{Theory of Neuromorphic Computing by Waves: Machine Learning by Rogue Waves, Dispersive Shocks, and Solitons},
    Phys. Rev. Lett. Vol. 125(9), 093901, 2020,
\url{https://link.aps.org/doi/10.1103/PhysRevLett.125.093901}.

\bibitem{Bile2023}
    A. Bile, H. Tari, R. Pepino, Riccardo, A. Nabizada, E. Fazio,
    \textit{Solitonic Neural Network: A novel approach of Photonic Artificial Intelligence based on photorefractive solitonic waveguides},
    EPJ Web Conf. Vol. 287, p. 13003, 2023,
    \url{https://doi.org/10.1051/epjconf/202328713003}.

\bibitem{Agrawal2019}
    G. P. Agrawal, 
    \textit{Nonlinear Fiber Optics},
    Academic Press,
    6th edition,
    2019.

\bibitem{Saeed2025}
    S. Saeed, M. M\"uft\"uoglu, G. R. Cheeran, T. Bocklitz, B. Fischer, M. Chemnitz,
    \textit{Nonlinear Inference Capacity of Fiber-Optical Extreme Learning Machines},
    pre-print, ArXiv, 2025,
    \url{https://arxiv.org/abs/2501.18894}.

\bibitem{Ermolaev2025}
    A. V. Ermolaev, M. Hary, L. Leybov, P. Ryczkowski, A. Skalli, D. Brunner, G. Genty, J. M. Dudley,
    \textit{Limits of nonlinear and dispersive Fiber Propagation for Photonic Extreme Learning},
    Opt. Lett. 50, 4166-4169, 2025,
    \url{https://doi.org/10.1364/OL.562186}.

\bibitem{Sozos2024}
  K. Sozos, S. Deligiannidis, C. Mesaritakis, A. Bogris,
  \textit{Unconventional Computing Based on Four Wave Mixing in Highly Nonlinear Waveguides.} 
  IEEE J. Quantum Electron. Vol. 60(4), pp. 1-6, 2024,
  \url{https://ieeexplore.ieee.org/document/10539121}.

\bibitem{Finot2015}
    C. Finot, 
    \textit{40-GHz photonic waveform generator by linear shaping of four spectral sidebands},
    Opt. Lett. Vol. 40(7), pp. 1422--1425, 2015,
    \url{https://doi.org/10.1364/OL.40.001422}.

\bibitem{Dudley2009}
    J. M. Dudley, G. Genty, F. Dias, B. Kibler, N. Akhmediev,
    \textit{Modulation instability, Akhmediev Breathers and continuous wave supercontinuum generation},
    Opt. Express Vol. 17(24), 2009,
    \url{https://opg.optica.org/oe/abstract.cfm?URI=oe-17-24-21497}.
    
\bibitem{Frisquet2013}
  B. Frisquet, B. Kibler, G. Millot,
  \textit{Collision of Akhmediev Breathers in Nonlinear Fiber Optics},
  Phys. Rev. X Vol. 3(4), 041032, 2013,
  \url{https://link.aps.org/doi/10.1103/PhysRevX.3.041032}.

\bibitem{Xu2019}
  G. Xu, A. Gelash, A. Chabchoub, V. Zakharov, B. Kibler, 
  \textit{Breather Wave Molecules},
  Phys. Rev. Lett. Vol. 122(8), 084101, 2019,
  \url{https://link.aps.org/doi/10.1103/PhysRevLett.122.084101}. 

\bibitem{Andral2020}
    U. Andral, B. Kibler, J. M. Dudley, C. Finot,
    \textit{Akhmediev breather signatures from dispersive propagation of a periodically phase-modulated continuous wave},
    Wave Motion Vol. 95, 102545, 2020, \url{https://www.sciencedirect.com/science/article/pii/S016521251930410X}.

\bibitem{Schiek2021}
    R. Schiek,
    \textit{Excitation of nonlinear beams: from the linear Talbot effect through modulation instability to Akhmediev breathers},
    Opt. Express 29(10), pp. 15830--15851, 2021,
    \url{https://opg.optica.org/oe/abstract.cfm?URI=oe-29-10-15830}.

\bibitem{Zajnulina2024}
    M. Zajnulina, M. B\"{o}hm,
    \textit{Temporal Talbot effect: from a quasi-linear Talbot carpet to soliton crystals and Talbot solitons},
    Opt. Lett. 49(14), pp. 3894--3897, 2024,
    \url{https://opg.optica.org/ol/abstract.cfm?URI=ol-49-14-3894}.

\bibitem{Suret2024}
    P. Suret, Pierre, S. Randoux, A. Gelash, D. Agafontsev, B. Doyon, G. El,
    \textit{Soliton gas: Theory, numerics, and experiments},
    Phys. Rev. E Vol. 109(6), 061001, 2024,
    \url{https://link.aps.org/doi/10.1103/PhysRevE.109.061001}.

\bibitem{Yamano2024}
    T. Yamano,
    \textit{Shannon entropy and fisher information of solitons for the cubic nonlinear Schr{\"o}dinger equation.}
    Eur. Phys. J. Plus Vol. 139, p. 595, 2024,
    \url{https://doi.org/10.1140/epjp/s13360-024-05402-w}.

\bibitem{Fisher1936}
    R. A. Fisher,
    \textit{The Use of Multiple Measurements in Taxonomic Problems.}
    Ann. Eugen. Vol. 7(2), pp. 179-188, 1936,
    \url{https://doi.org/10.1111/j.1469-1809.1936.tb02137.x}.

\bibitem{Wisconsin1993}
    W. Wolberg, O. Mangasarian, N. Street, W. Street,
    \textit{Breast Cancer Wisconsin (Diagnostic)}, Dataset, 1993.
    UCI Machine Learning Repository 
    \url{https://doi.org/10.24432/C5DW2B}.

\bibitem{Boehm_2006} 
  M. B\"ohm, F. Mitschke,
  \textit{Soliton-radiation beat analysis},
  Phys. Rev. E Vol. 73(6), 066615, 2006,
  \url{https://link.aps.org/doi/10.1103/PhysRevE.73.066615}.

\bibitem{Zajnulina_2015}
    M. Zajnulina, M. B\"ohm, K. Blow, A. A. Rieznik, D. Giannone, R. Haynes, M. M. Roth,
    \textit{Soliton radiation beat analysis of optical pulses generated from two continuous-wave lasers},
    Chaos Vol. 25(10), 2015,
    \url{https://doi.org/10.1063/1.4930316}.

\bibitem{Zajnulina_2017}
    M. Zajnulina, M. B\"ohm, D. Bodenm\"uller, K. Blow, J.M. Chavez Boggio, A.A. Rieznik, M.M. Roth,
    \textit{Characteristics and stability of soliton crystals in optical fibres for the purpose of optical frequency comb generation},
    Opt. Commun. Vol. 393, pp. 95-102, 2017,
    \url{https://doi.org/10.1016/j.optcom.2017.02.035}.

\bibitem{Mitschke2017}
    F. Mitschke, C. Mahnke, A. Hause, 
    \textit{Soliton Content of Fiber-Optic Light Pulses}, 
    Appl. Sci. Vol. 7(6), 635, 2017, \url{https://doi.org/10.3390/app7060635}.
    
\bibitem{Cincotta2021}
    P. M. Cincotta, C. M. Giordano, R. Alves Silva, C. Beaug\'e,
    \textit{The Shannon entropy: An efficient indicator of dynamical stability},
    Phys. D: Nonlinear Phenom. Vol. 417, p. 132816, 2021.
    \url{https://doi.org/10.1016/j.physd.2020.132816}.

\bibitem{Duran2022}
    A. Lozano-Dur\'an, Adri\'an, G. Arranz,
    \textit{Information-theoretic formulation of dynamical systems: Causality, modeling, and control},
    Phys. Rev. Res. Vol. 4(2), p. 023195, 2022,
    \url{https://doi.org/10.1103/PhysRevResearch.4.023195}.

\bibitem{Kuyper1996}
    I. G. Sprinkhuizen-Kuyper,
    \textit{Some remarks on the entropy of a neural network},
    BENELEARN-96, 1996.

\bibitem{LIMA2022}
    F.C.E. Lima,
    \textit{Quantum information entropies for a soliton at hyperbolic well},
    Ann. Phys. Vol. 442, p. 168906, 2022,
    \url{https://doi.org/10.1016/j.aop.2022.168906}.

\bibitem{Hult2007}
    J. Hult,
    \textit{A Fourth-Order Runge–Kutta in the Interaction Picture Method for Simulating Supercontinuum Generation in Optical Fibers.}
    J. Light. Technol. Vol. 25(12), pp. 3770-3775, 2007,
    \url{https://doi.org/10.1109/JLT.2007.909373}.




\bibitem{scikit-learn}
    F. Pedregosa, G. Varoquaux, A. Gramfort, V. Michel,
          B. Thirion, O. Grisel, M. Blondel, P. Prettenhofer,
          R. Weiss, V. Dubourg, J. Vanderplas, A. Passos, D
          Cournapeau, M. Brucher, M. Perrot, E. Duchesnay,
    \textit{Scikit-learn: Machine Learning in Python},
    JMLR Vol. 12, pp. 2825--2830, 2011.


\bibitem{Jannson_1981}
    T. Jannson, J. Jannson,
    \textit{Temporal self-imaging effect in single-mode fibers},
    J. Opt. Soc. Am. Vol. 71(11), pp. 1373--1376, 1981,
    \url{https://opg.optica.org/abstract.cfm?URI=josa-71-11-1373}.

\bibitem{Wu_2023}
    J. Wu, M. Clementi, E. Nitiss, J. Hu, C. Lafforgue, C.-S. Br{\`e}s,
    \textit{Bright and dark Talbot pulse trains on a chip},
    Commun. Phys. Vol. 6(1), pp. 2399-3650, 2023,
    \url{https://doi.org/10.1038/s42005-023-01375-x}.

\bibitem{Kimmoun_2016}
    O. Kimmoun, H. C. Hsu, H. Branger, M. S. Li, Y. Y. Chen, C. Kharif, M. Onorato, E. J. R. Kelleher, B. Kibler, N. Akhmediev, A. Chabchoub,
    \textit{Modulation Instability and Phase-Shifted Fermi-Pasta-Ulam Recurrence},
    Sci. Rep. 6, 28516, 2016,
    \url{https://doi.org/10.1038/srep28516}.

\bibitem{Hammani_2011}
    K. Hammani, B. Kibler, C. Finot, P. Morin, J. Fatome, J. M. Dudley, G. Millot, 
    \textit{Peregrine soliton generation and breakup in standard telecommunications fiber}, 
    Opt. Lett. 36, pp. 112-114, 2011,
    \url{https://doi.org/10.1364/OL.36.000112}.

\bibitem{Ye_2020}
    Y. Ye, L. Bu, W. Wang, S. Chen, F. Baronio, D. Mihalache,
    \textit{Peregrine Solitons on a Periodic Background in the Vector Cubic-Quintic Nonlinear Schrödinger Equation},
    Front. Phys. 8, 2020, \url{https://www.frontiersin.org/journals/physics/articles/10.3389/fphy.2020.596950}.

\bibitem{Cohen_2008}
    O. Cohen, L. Chong, T. Schwartz, T. Popmintchev, M. M. Murnane, H. C. Kapteyn,
    \textit{Talbot solitons,}
    Opt. Lett. Vol. 33(8), pp. 830--832, 2008,
    \url{https://opg.optica.org/ol/abstract.cfm?URI=ol-33-8-830}.

\bibitem{Wen_2011}
    J. Wen, Y. Zhang, S.-N. Zhu, M. Xiao,
    \textit{Theory of nonlinear Talbot effect,}
    J. Opt. Soc. Am. B Vol. 28(2), pp. 275--280, 2011,
    \url{https://opg.optica.org/josab/abstract.cfm?URI=josab-28-2-275}.

\bibitem{Zhang_2014}
  Y. Zhang, M. R. Beli\'{c}, H. Zheng, H. Chen, C. Li, J. Song, Y. Zhang,
  \textit{Nonlinear Talbot effect of rogue waves},
  Phys. Rev. E Vol. 39(3), 032902, 2014,
  \url{https://link.aps.org/doi/10.1103/PhysRevE.89.032902}.

\bibitem{Naji_2021}
    M. A. Naji, S. El Filali, K. Aarika, EL H. Benlahmar, R. Ait Abdelouhahid, O. Debauche,
    \textit{Machine Learning Algorithms For Breast Cancer Prediction And Diagnosis},
    Procedia Comput. Sci. Vol. 191, pp. 487--492, 2021,
    \url{https://doi.org/10.1016/j.procs.2021.07.062}.

\bibitem{Morison_2024}
  H. Morison, J. Singh, N. Al Kayed, A. Aadhi, M. Moridsadat, M. Tamura, A. N. Tait, B. Shastri,
  \textit{Nonlinear dynamics in neuromorphic photonic networks: Physical simulation in Verilog-A},
  Phys. Rev. Appl. Vol. 21(3), p. 034013, 2024,
  \url{https://link.aps.org/doi/10.1103/PhysRevApplied.21.034013}.

\bibitem{Sunada_2019}
    S. Sunada, A. Uchida,
    \textit{Photonic reservoir computing based on nonlinear wave dynamics at microscale}
    Sci. Rep. 9, 19078, 2019,
    \url{https://doi.org/10.1038/s41598-019-55247-y}.

\bibitem{Kesgin_2025}
    B. U. Kesgin, U. Teğin,
    \textit{Photonic neural networks at the edge of spatiotemporal chaos in multimode fibers} 
    Nanophotonics, 2025,
    \url{https://doi.org/10.1515/nanoph-2024-0593}.

  
\end{thebibliography}



\end{document}